\documentclass[aps,pretwocolumn,amsmath,amssymb,superscriptaddress,reprint,longbibliography]{revtex4-1}
\usepackage{amsfonts,amsmath,amssymb,bm}
\usepackage{graphicx,graphics,float}
\usepackage{bm}
\usepackage{amssymb}
\usepackage{colordvi}
\usepackage{graphicx}
\usepackage{color}
\usepackage[colorlinks=true,linkcolor=blue,citecolor=blue,urlcolor=blue]{hyperref}%
\usepackage{hyperref}
\usepackage{comment}
\usepackage{harpoon}

\newcommand{\ov}[1]{\overline{{#1}}}
\newcommand{\be}{\begin{equation}}
\newcommand{\ee}{\end{equation}}
\newcommand{\bea}{\begin{eqnarray}}
\newcommand{\eea}{\end{eqnarray}}

\newcommand{\vep}{\varepsilon}
\newcommand{\ave}[1]{\langle #1\rangle}
\newcommand{\ome}{\omega}

\def\nn{\nonumber}

\begin{document}

\title{Inelastic thermoelectric transport and fluctuations in mesoscopic system}

\author{Rongqian Wang}
\affiliation{Institute of Theoretical and Applied Physics, School of Physical Science and Technology \&
Collaborative Innovation Center of Suzhou Nano Science and Technology, Soochow University, Suzhou 215006, China.}

\author{Chen Wang}\email{wangchen@zjnu.cn}
\address{Department of Physics, Zhejiang Normal University, Jinhua, Zhejiang 321004, China}

\author{Jincheng Lu}\email{jincheng.lu1993@gmail.com}
\address{Jiangsu Key Laboratory of Micro and Nano Heat Fluid Flow Technology and Energy Application, School of Physical Science and Technology, Suzhou University of Science and Technology, Suzhou, 215009, China}

\affiliation{Institute of Theoretical and Applied Physics, School of Physical Science and Technology \&
Collaborative Innovation Center of Suzhou Nano Science and Technology, Soochow University, Suzhou 215006, China.}

\address{Center for Phononics and Thermal Energy Science, China-EU Joint Center for Nanophononics, Shanghai Key Laboratory of Special Artificial Microstructure Materials and Technology, School of Physics Science and Engineering, Tongji University, Shanghai 200092 China}

\author{Jian-Hua Jiang}\email{jianhuajiang@suda.edu.cn}
\affiliation{Institute of Theoretical and Applied Physics, School of Physical Science and Technology \&
Collaborative Innovation Center of Suzhou Nano Science and Technology, Soochow University, Suzhou 215006, China.}

\date{\today}
\begin{abstract}
In the past decade, a new research frontier emerges at the interface between physics and renewable energy, termed as the inelastic thermoelectric effects where inelastic transport processes play a key role. The study of inelastic thermoelectric effects broadens our understanding of thermoelectric phenomena and provides new routes towards high-performance thermoelectric energy conversion. Here, we review the main progress in this field, with a particular focus on inelastic thermoelectric effects induced by the electron-phonon and electron-photon interactions. We introduce the motivations, the basic pictures, and prototype models, as well as the unconventional effects induced by inelastic thermoelectric transport. These unconventional effects include the separation of heat and charge transport, the cooling by heating effect, the linear thermal transistor effect, nonlinear enhancement of performance, Maxwell demons, and cooperative effects. We find that elastic and inelastic thermoelectric effects are described by significantly different microscopic mechanisms and belong to distinct linear thermodynamic classes. We also pay special attention to the unique aspect of fluctuations in small mesoscopic thermoelectric systems. Finally, we discuss the challenges and future opportunities in the field of inelastic thermoelectrics.
\end{abstract}

\maketitle

\tableofcontents

\section{Introduction}
\begin{figure}[htb]
\begin{center}
\centering\includegraphics[width=8.5cm]{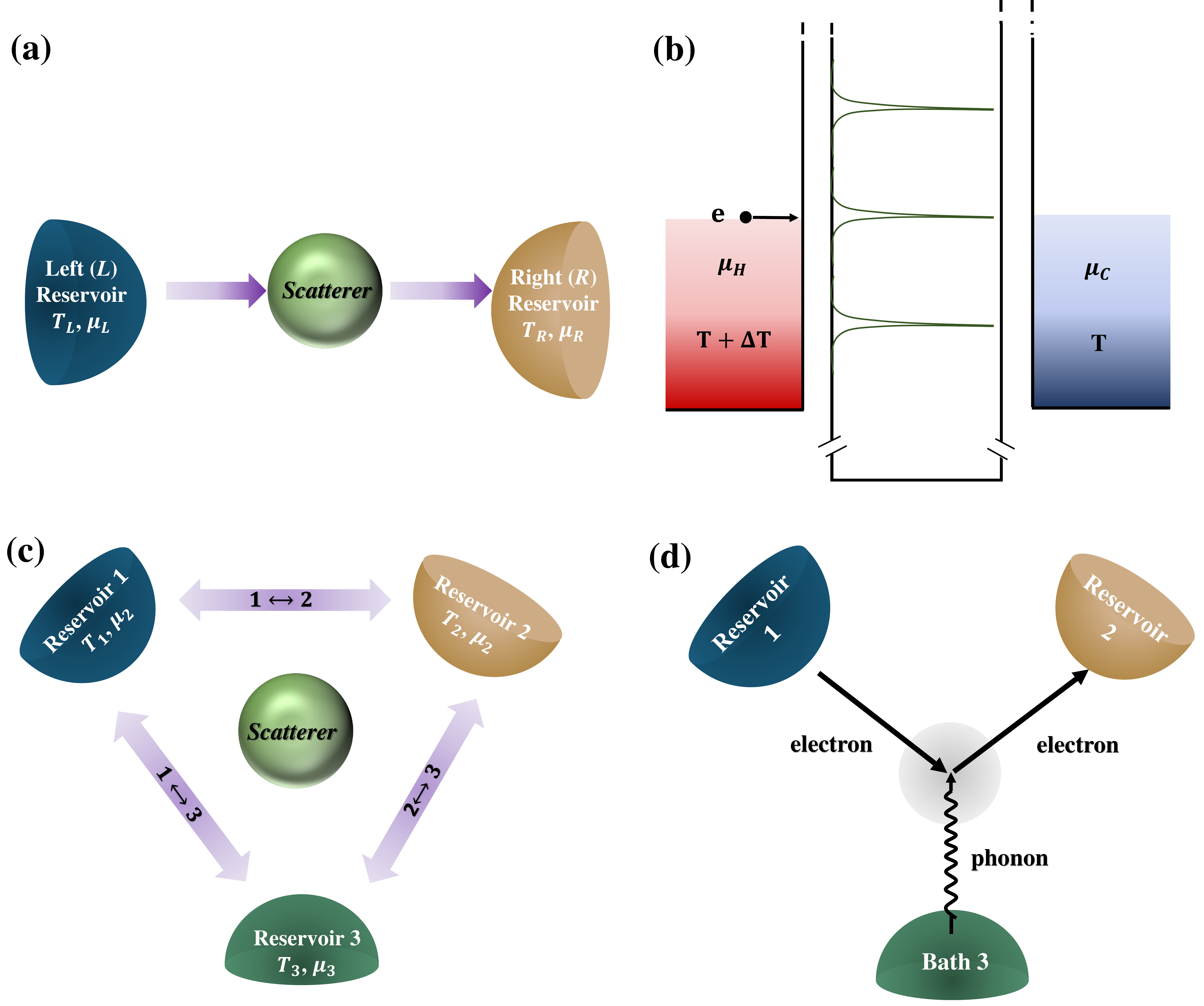}
\caption{(a) Schematic of Landauer's theory of transport between two reservoirs. (b) A typical system that can be described by Landauer's transport theory: resonant tunneling through quantum dots. (c) B\"uttiker's theory of decomposing complex transport in multi-terminal systems into many pairs of two-terminal transport processes. (d) An inelastic transport process that cannot be decomposed into pairs of two-terminal transport processes.}
\label{fig:sketch}
\end{center}
\end{figure}

The study of the fundamental science of thermoelectric effects will inevitably encounter the tight connection between quantum physics and thermodynamics when describing the microscopic processes~\cite{harman,yimry1997book}. Starting from the 1980's when the fundamental theory of mesoscopic transport was applied to investigate thermoelectric transport [see illustrations in Figs.~\ref{fig:sketch}(a) and~\ref{fig:sketch}(b)], the basic elements of quantum mechanics appear in this field, e.g, coherent transport~\cite{BrandnerCoherent,BrandnerPRL17,SanchezPRB21,BrandnerPRX21}, dephasing and dissipation~\cite{LeggettRMP,PED}, Onsager's reciprocal relationship~\cite{Onsager1,Onsager2,CallenPRB}, and broken time-reversal symmetry induced thermodynamic bounds~\cite{BRTS4,Saito2011,JiangPRE}. Moreover, the quantum confinement effect efficiently tunes the density-of-states of electrons, and thus considerably modifies the thermoelectric performance using nano- and mesostructures. However, for decades long, the focus is mainly on elastic transport that can be decomposed via B\"utikker's theory of multi-terminal transport into many pairs of two-terminal processes, which is commonly believed to be sufficient to describe all thermoelectric transport phenomena~\cite{jauho2008book,Buttiker1,Buttiker-4T,buttiker1987,WingreenPRL,WingreenPRB94,ymblanter2000pr,WangPRB06,LvPRB07,wang08,Wang2014,lvPRB16,LvPRB17,BrandnerPRL2018,Brandner4T,tu21,CarregaPRXQ}.

Starting from a decade ago, another \textcolor{blue}{fundamental} category of nonequilibrium processes, i.e., the inelastic transport processes, have attracted increasing attention~\cite{guo26,Nanotechnology,JiangCRP,ThierschmannCRP,OraPRB2010,Lena2012,ArracheaPRB14,RoyATS,ZhouPRB15,HenrietPRB15,McConnell22}.
The community gradually became aware of its oddness, which is essentially due to that these inelastic processes cannot be decomposed into many pairs of two-terminal processes [see illustration in Fig.~\ref{fig:sketch}(c)]. Note that strictly speaking, there exist inelastic processes that can be decomposed in such a way, \textcolor{blue}{which is not within the scope of this review}. For instance, the Mott-Cutler theory of thermoelectric transport can include inelastic processes between two terminals~\cite{PRB69}. The main focus of this review is on the inelastic processes that cannot be described by B\"utikker's theory. In fact, in many cases, the basic structure of such inelastic processes is the correlated transport among \textcolor{blue}{multiple} terminals \textcolor{blue}{at the quantum mechanical level}, as illustrated in Fig.~\ref{fig:sketch}(d).

In the past decades, based on the elastic transport theory, many efforts have been devoted to improving the thermoelectric figure of merit $ZT$ by investigating and engineering the microscopic transport mechanisms to achieve enhanced the electrical conductivity and reduced the thermal conductivity~\cite{HicksPRB1,HicksPRB2,science99,VenPRB,science08,2008Complex,Biswas2012,ZhaoLDScience}. However, the inevitible correlation between charge and heat transport in the conventional Mott-Cutler theory for thermoelectric effect sets a bottleneck to such an approach. Here, we focus on an alternative approach where the thermoelectric effect is induced instead by inelastic transport mechanisms. In this regime, the theory of thermoelectric figure of merit $ZT$ must be reconsidered. Indeed, the figure of merit, the optimal energy efficiency and output power for inelastic thermoelectric effect are quite different from their conventional counterparts~\cite{Jiang2012,JiangNJP,JiangJAP,JiangCRP,Jiang2017,Rongqian,JiangNearfield,wangPRApplied,MyPRBdemon}. Research in this direction indicates that inelastic thermoelectric effects could be a promising approach towards next-generation high-performance thermoelectric energy conversion and functional devices~\cite{JiangCRP,Nanotechnology,Jiangtransistors}.

\begin{figure}[htb]
\begin{center}
\centering\includegraphics[width=8.0cm]{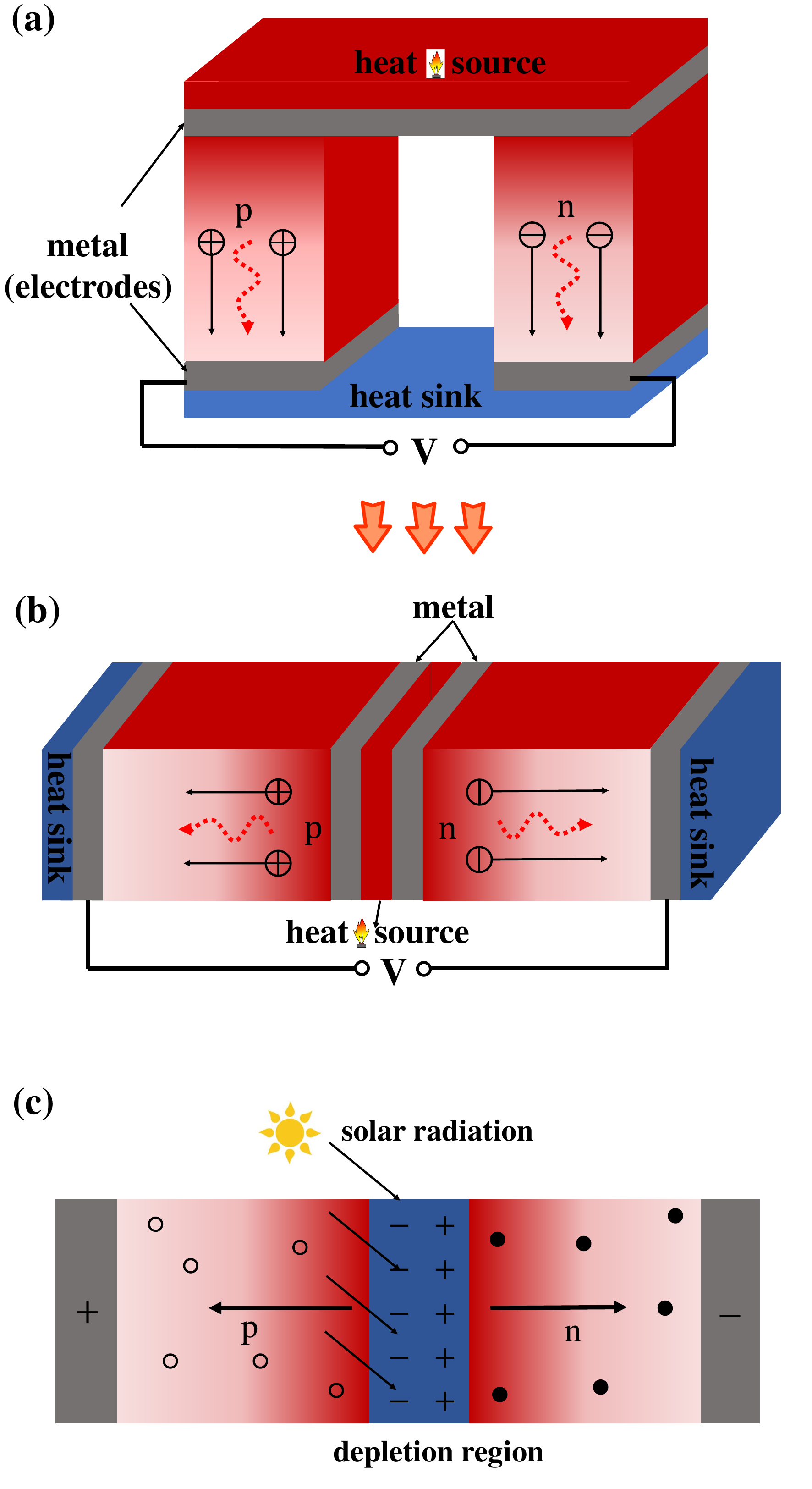}
\caption{(a) Schematic of the conventional thermoelectric energy harvester that converts heat to electricity. It is composed of two electrically conducting materials: one $n$-type and the other $p$-type. They are joined at the top by a metal (electrode) to make a junction.  When the junction is heated, both types of carriers conduct heat to the cold base and a voltage bias is generated at the two base electrodes. (b) The unfolded geometry of the thermoelectric energy harvesting device. (c) Schematic of a solar cell based on a $p$-$n$ junction. Figures (a) and (b) are reproduced from Ref.~\cite{JiangSR}.}\label{fig:p-n-juction}
\end{center}
\end{figure}

\textcolor{blue}{Before elaborating on the various surprising properties, we first give an inspirative comparison between thermoelectric engine and solar cells. As shown in Fig.~\ref{fig:p-n-juction}(a), a conventional thermoelectric engine consists of two types of semiconductor materials. One of them is $n$-doped, while the other is $p$-doped. Their electrical connection and thermal contact with the heat source and sink are achieved in a bridge like structure in the figure. If this structure is stretched to be straight, as shown in Fig.~\ref{fig:p-n-juction}(b), it becomes similar to a solar cell [Fig.~\ref{fig:p-n-juction}(c)]. An interesting question arises: why solar cells are more efficient than thermoelectric heat engines, despite that their structures are similar? The key difference between solar cells and thermoelectric heat engines are that they rely on different transport mechanisms. Thermoelectric heat engines rely on diffusive thermoelectric transport in the $p$- and $n$-types of semiconductors which are connected by Ohmic contact via metal electrodes. In contrast, solar cells rely on photo-carrier generation and carrier splitting due to the built-in electric field in the depletion region of the $p$-$n$ junction. While the diffusive thermoelectric transport is based mostly on the elastic transport processes, the photo-carrier generation due to solar radiation is typical inelastic transport processes in semiconductors. Another significant difference is that there are three reservoirs in solar cells, the source, the drain and the Sun. Energy exchange simultaneously takes place among these resoirs. In contrast, there are only two reservoirs in a thermoelectric heat engine. We believe that the much higher energy efficiency in solar cells (typically $>20\%\eta_C$ with $\eta_C=1-T_E/T_S$ being the Carnot efficiency of solar cells where $T_E$ is the ambient temperature on earth and $T_S$ is the black-body radiation temperature of the Sun)~\cite{shockley1961detailed,Photocell}, as compared with the lower energy efficiency of thermoelectric heat engines (typically $<20\%\eta_C^\prime$ with $\eta_C^\prime=1-T_c/T_h$ denoting the Carnot efficiency of thermoelectric heat engines where $T_c$ and $T_h$ are the temperatures of the cold and hot reservoirs)~\cite{Jaziri20,Tohidi} is not only due to their differences in the Carnot efficiency, $\eta_C\gg \eta_C^\prime$, but also due to the above differences in their transport mechanisms and thermodynamic properties. These differences may also be responsible for the much higher output power in solar cells~\cite{JiangNJP,Jiang2017,Rongqian}. The above thinking inspired us to study inelastic thermoelectric transport in the aim of developing an approach toward high-performance thermoelectric energy conversion beyond the conventional one. In this sense, solar cells are a particular type of inelastic thermoelectric systems that have been put into industrial applications. Solar cells also provide a prototype demonstration on how mesoscopic inelastic thermoelectric systems can be integrated into macroscopic devices. This review is dedicated to the efforts devoted to the emergent field of inelastic thermoelectric effects which seeks for deeper understanding of the underlying physics, generalization of the physical mechanisms, exploration of new effects and new material systems, and investigation of new applications.}

\textcolor{blue}{In the past decade, research on inelastic thermoelectric effects has made notable progresses. There are different types of inelastic thermoelectric systems. For instance, phonon-assisted inelastic thermoelectric systems~\cite{Jiang2012,Jiang2013,JiangNJP,JiangCRP,Jiangtransistors,Jiang2017,BijayPRB16,BijayPRBTunable,BijayPRB19}, photon-assisted inelastic thermoelectric systems~\cite{CleurenPRB09,Cooling2,JiangNearfield,wangPRApplied,houckNP}, magnon-assisted inelastic thermoelectric systems~\cite{Sothmann2013}, and those systems where inelastic transport is assisted by Coulomb interactions between electrons~\cite{Rafael,SanchezNJP13,SothmannPRB12,WhitneyPhysE,ChenJC1,ChenJC2,MayrhoferPRB21}. Due to the limited space, we focus in this review phonon- and photon-assisted inelastic thermoelectric effects. In this context, we point out that Coulomb-assisted inelastic thermoelectric systems were reviewed in Ref.~\cite{Nanotechnology}. At this point, it is necessary to state that to have the inelastic thermoelectric effects well-defined, we need the phonons (or other collective excitations) to have a temperature different from the electrons. This condition often cannot be met in macroscopic systems, therefore we discuss inelastic thermoelectric effects mainly in mesoscopic systems. However, it is possible, via micro fabrication technologies, to integrate these mesoscopic inelastic thermoelectric systems into macroscopic devices. As stated above, solar cells are successful demonstration of such integration.}

\textcolor{blue}{In this review, we start with a general analysis of thermoelectric transport in mesoscopic systems where elastic and inelastic transport processes are formulated with equal footing. Based on this, we give the bounds on linear transport coefficients for the elastic and inelastic transport processes, respectively. In Sec.~\ref{DQD-device}, we discuss a simple model for phonon-assisted inelastic thermoelectric transport. The unconventional thermoelectric effects induced by the inelastic transport such as rectification, transistor, cooling by heating, and cooling by thermal current effects in the nonlinear regime are considered in Sec.~\ref{inelastic-effects}. Effects that can lead to enhancement of thermoelectric performance, such as the nonlinear transport effect, cooperative effect, and near-field effect are also introduced. The statistics of efficiency for three-terminal systems with (broken) time-reversal symmetry, the thermal transistor amplification factor, and the cooling by heating energy efficiency under the Gaussian fluctuation framework are reviewed in Sec.~\ref{statistics}. Thermophotovoltaic systems with near-field enhancement are also reviewed as a special category of inelastic thermoelectric systems. Finally, we summarize and give outlooks in Sec.~\ref{conclusion}.}

\section{Elastic versus inelastic thermoelectric transport in mesoscopic systems}

\textcolor{blue}{Thermoelectric transport in mesoscopic systems is driven by thermodynamics forces (e.g., temperature gradients and voltage biases). The steady-state transport is characterized by electrical currents and heat currents. The latter consists of contributions from electrons and other quasiparticles such as phonons and photons. Thermoelectric transport can generally be categorized into two main classes: i) elastic transport and ii) inelastic transport~\cite{Jiangtransistors,MyPRBtransistor}.}

When a mesoscopic system is connected with two electronic reservoirs, the voltage bias $V$ and the temperature difference $\Delta T = T_h - T_c$ between the two reservoirs (hot and cold reservoirs, with temperatures $T_h > T_c$) drive a charge current $I_e$ and a heat current $I_Q$. In the linear-response regime, the charge and heat currents are related to the thermodynamic affinities (i.e., the voltage bias and the temperature difference) via the Onsager matrix~\cite{JiangOra}
\begin{equation}
\begin{aligned}
\left( \begin{array}{cccc} I_e\\ I_Q \end{array}\right) =
 \left( \begin{array}{cccc} G & L \\ L &
    K \end{array} \right) \left( \begin{array}{cccc}  V
    \\ \Delta T/T \end{array}\right),
\label{eq:2OnsagerMatrix}
\end{aligned}
\end{equation}
which is time-reversal symmetry~\cite{Onsager1,Onsager2}. $T$ is the average temperature of the system. $G$ and $K$ denote the charge and heat conductivities, respectively. $L$ represents thermoelectric effect and the thermopower (or Seebeck coefficient) is $S=L/(TG)$~\cite{Yamamoto17PRB}. The energy efficiency of the two-terminal thermoelectric system is limited by the second law of thermodynamics~\cite{PRXQuantum}. In the linear-response regime the maximum efficiency is given by~\cite{harman,yimry1997book,gchen2005book,jauho2008book}
\begin{equation}
\eta_{\rm max} = \eta_C\frac{\sqrt{1+ZT} - 1}{\sqrt{1+ZT} + 1}\le \eta_C,
\end{equation}
where $\eta_C=1-T_c/T_h$ is the Carnot efficiency. The maximum efficiency show monotonous increase as a function of the dimensionless figure of merit $ZT$, where $ZT=TG S^2/K$. Clearly, the maximum efficiency $\eta_{\rm max}$ approaches the Carnot efficiency $\eta_C$ when $ZT$ approaches $\infty$. Unfortunately, high values of $ZT$ are difficult to be achieved. In the definition of $ZT$, the heat conductivity $K$ consists of both the electronic heat conductivity and the phononic heat conductivity. In particular, Mahan and Sofo proposed that the ``best thermoelectrics'' can be realized in narrow-band conductors~\cite{Mahan}. Their proposal is based on the arguments that electronic heat conductivity can be suppressed in these narrow-band conductors, while a decent Seebeck coefficient can still be achieved. {\color{blue} However, this argument leads to a lot of debates~\cite{ZhouPRL}, and phonon thermal transport will inevitably suppress the figure of merit in these narrow-band conductors.} This reveals that there exists intrinsic correlation between the charge and heat transport, since they are both carried by electrons. Though the separation of the charge and heat transport is impossible in elastic transport processes, we will show that it becomes possible in inelastic transport processes.

\subsection{The elastic thermoelectric transport: from two-terminal to multiple-terminal setup}

{\color{blue}Landauer's scattering theory is an effective description of quantum transport in a two-terminal setup}~\cite{Landauer57,Landauer70,MazzaNJP}, as shown in Fig.~\ref{fig:sketch}(a). Latter, B\"{u}ttiker's multi-terminal version of the scattering theory was placed on a more solid theoretical footing by Ref.~\cite{Aaron}, which derived it from the Kubo linear-response formalism [see Fig.~\ref{fig:sketch}(c)]. The Landauer-B\"{u}ttiker scattering theory is capable of describing the electrical, thermal, and thermoelectric properties of non-interacting electrons in an arbitrary potential, in terms of the probability that the electrons go from one reservoir to another.

Moreover, the Landauer-B\"{u}ttiker scattering theory is {\it only}  applicable to ``elastic transport process'',
with each microscopic process only involving two reservoirs.  Based on the standard Landauer-B\"{u}ttiker theory~\cite{Sivan,butcher1990,BENENTI20171}, the elastic electronic currents are expressed as
\begin{equation}
\begin{aligned}
&I_e^i|_{\rm el}=\frac{e}{h}\int_{-\infty}^{-\infty}dE\sum_{i\ne j}{\mathcal{T}_{i\rightarrow j}[f_i(E) -f_j(E)]},\\
&I_{Q}^i|_{\rm el}=\frac{1}{h}\int_{-\infty}^{-\infty}dE\sum_{i\ne j}{\mathcal{T}_{i\rightarrow j}(E-\mu)[f_i(E) -f_j(E)]},
\end{aligned}\label{eq:JmIm}
\end{equation}
respectively, where $\mathcal{T}_{ij}$ is the transmission function from reservoir $j$ to reservoir $i$, $f_i=\{\exp[(E-\mu_i)/k_BT_i]+1\}^{-1}$ is the Fermi-Dirac distribution function, with temperature $T_i$ in the {\it{i}}th fermion bath and $\mu_i$ the corresponding chemical potential. $k_B$ is the Boltzmann constant. Moreover, the probability conservation requires that $\sum_{ij}\mathcal{T}_{i\rightarrow j}=1$~\cite{JiangCRP}. From Eq.~\eqref{eq:JmIm}, it is known that elastic currents are dominated by the two-terminal nonequilibrium processes.

\begin{figure}[htb]
\begin{center}
\centering\includegraphics[width=7.5cm]{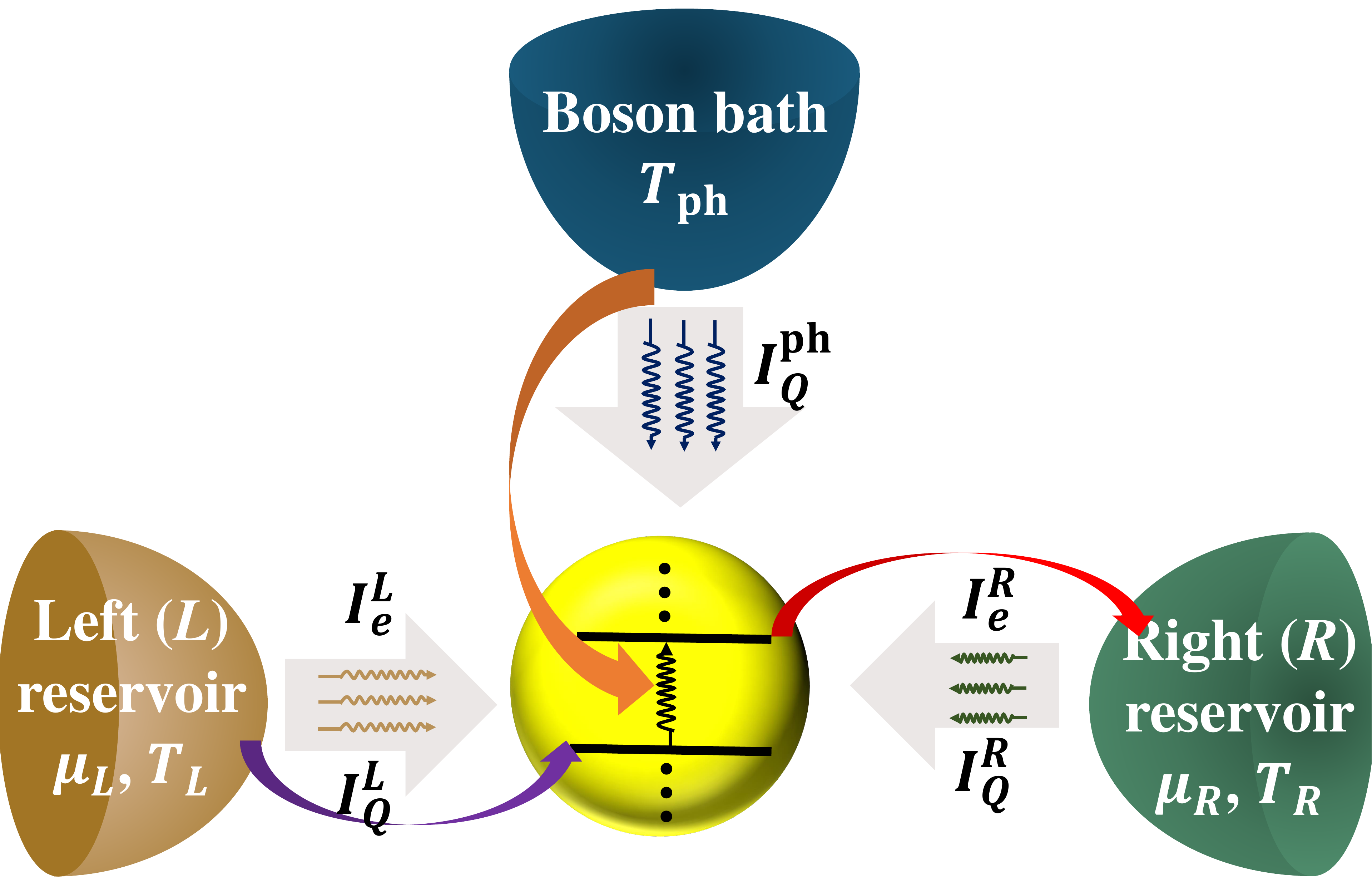}
\caption{Illustration of possible boson-assisted inelastic transport processes. $I_e^i$ ($i=L,R$) denoting the electronic current flowing from the $i$th reservoir and $I_Q^i$ ($i=L,R,{\rm ph}$) denoting the heat current flowing from the $i$th reservoir. Average currents $I_i$ are positive when flowing towards the system. }\label{inelastic-skeptch}
\end{center}
\end{figure}

\subsection{The inelastic thermoelectric transport assisted by a boson bath}

For the three-terminal setup shown in Fig.~\ref{inelastic-skeptch},
 the electronic (from reservoir $L$ and $R$) and bosonic heat currents (from boson bath) are nonlinearly coupled.
 Such nonlinearity mainly stems from the inelastic electron-phonon scattering process, which cooperatively involve three reservoirs.
 This phonon-assisted transport process in the three-terminal nanodevices is termed as ``inelastic transport process''. We emphasize that the inelastic transport process in this work is defined for reservoirs (terminals) rather not particles.
 Therefore, the inelastic transport processes must involve interactions between particles from at least three different terminals.
 While processes involving only two terminals, although they may involve interactions and energy exchange between quasiparticles, are still elastic transport processes. The inelastic transport process in the present work thus unveils a large number of processes ignored in the conventional study of the mesoscopic transport. It is interesting to note that the inelastic current densities flowing into these terminals are the same. Specifically, the inelastic heat currents flowing into the three reservoirs are expressed via the Fermi golden rule~\cite{Jiang2012,Jiang2013}
\begin{equation}
\begin{aligned}
&I_{e}^{L}|_{\rm inel}=-I_{e}^{R}|_{\rm inel}=\iint dE_1d\omega_3j_{\rm in}(E_{1},\omega_3),\\
&I_{Q}^{L}|_{\rm inel}=\iint dE_1d\omega_3(E_{1}-\mu_{1})j_{\rm in}(E_1,\omega_3),\\
&I_{Q}^{R}|_{\rm inel}=\iint dE_1d\omega_3(E_{1}-\mu_{2}+\omega_3)j_{\rm in}(E_1,\omega_3),\\
&I_{Q}^{\rm ph}|_{\rm inel}=\iint dE_1d\omega_3\omega_3j_{\rm in}(E_1,\omega_3),
\end{aligned}~\label{eq:J1J2I3}
\end{equation}
where $j_{\rm in}(E_1,\omega_3) = C_{\rm in}f_1(E_1)[1-f_2(E_2)]N_B(\omega_3)-C_{in}f_2(E_2)[1-f_1(E_1)][1+N_B(\omega_3)]$, $E_2-E_1=\omega_3$ and $N_B(\omega)=[\exp(\omega/k_BT_i)-1]^{-1}$ being the Bose-Einstein distribution function.
The transition coefficient $C_{\rm in}$ is the probability for electrons/bosons to tunnel from the $i$th reservoir into the scatterer.

\section{Bound on the linear transport coefficients for elastic and inelastic transport}

Typically, for a scatterer interacting with three reservoirs, we have three corresponding heat currents. However, due to the heat current conservation ($I_Q^L+I_Q^R+I_Q^{\rm ph}=0$) in the linear-response regime,  two of them are independent, e.g., $I_Q^L$ and $I_Q^{\rm ph}$ are two independent heat currents. The transport equation of these heat currents can be expressed as~\cite{JiangPRE}
\begin{equation}
\begin{aligned}
\left( \begin{array}{cccc} I_Q^L\\ I_Q^{\rm ph} \end{array}\right) =
 \left( \begin{array}{cccc} K_{11} & K_{12} \\ K_{12} &
    K_{22} \end{array} \right) \left( \begin{array}{cccc} \frac{T_L-T_R}{T_R}\\
    \frac{T_{\rm ph}-T_R}{T_R} \end{array}\right).
\label{eq:2OnsagerMatrix}
\end{aligned}
\end{equation}
$K_{11(22)}$ and $K_{12}$ are the diagonal and off-diagonal
thermal conductances, which were originally derived based
on the Onsager theory. These two coefficients are obtained by $K_{11}=\partial I_Q^L/\partial T_L$, $K_{12}=\partial I_Q^L/\partial T_{\rm ph}$, and $K_{22}=\partial I_Q^{\rm ph}/\partial T_{\rm ph}$ in the limit $T_L$, $T_R$, $T_{\rm ph}\rightarrow T$ with $|T_{L({\rm ph})}-T_R|\ll T_R$.

\begin{figure}[htb]
\begin{center}
\centering\includegraphics[width=5.0cm]{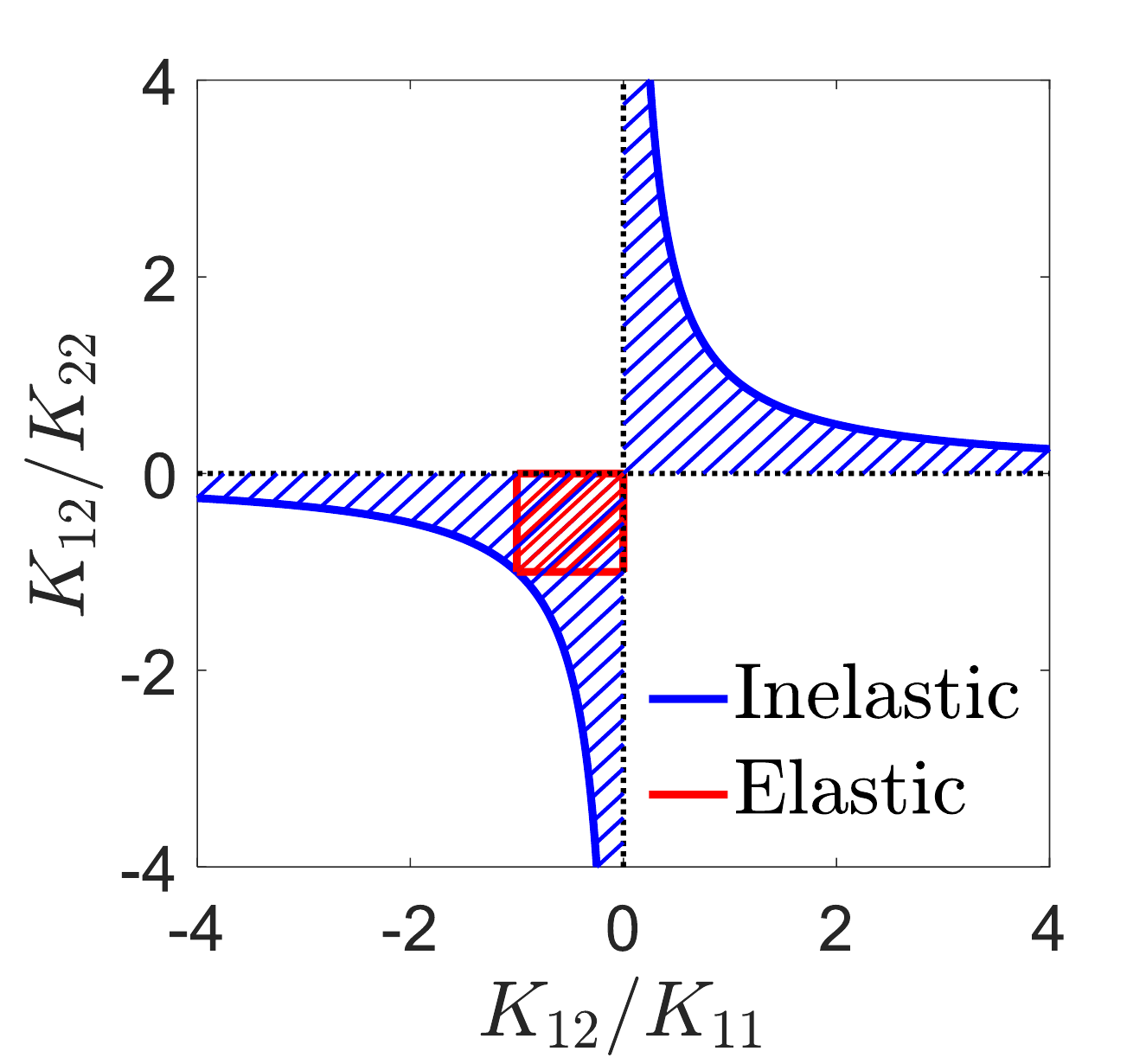}
\caption{ The boundary of the Onsager coefficients. The shaded blue area represents the broadening of the inelastic case, the shaded red area represents the broadening of the elastic case.}\label{fig:bound}
\end{center}
\end{figure}

Then, the bounds of Onsager coefficients based on elastic and inelastic scattering mechanisms can be described, separately. We first consider the generic elastic transport. The elastic coefficients are specified as
\begin{align}
K^{\rm el}_{ij}=\left \langle E^2\right \rangle_{ij} G^{\rm el}_{ij} \quad (i=1,2,3),
\end{align}
where the average under the elastic processes is given by~\cite{MyJAP}
\begin{equation}
\left \langle {\mathcal O(E)} \right \rangle_{ij}=\frac{\int{dE{\mathcal O(E)}G^{\rm el}_{ij}(E)}}{\int{dEG^{\rm el}_{ij}(E)}},
\end{equation}
with the probability weight
\begin{subequations}
\begin{align}
G^{\rm el}_{11}(E)&=({\mathcal T_{12}}+{\mathcal T_{13}})f(E)[1-f(E)],\\
G^{\rm el}_{12}(E)&=(-{\mathcal T_{13}})f(E)[1-f(E)],\\
G^{\rm el}_{22}(E)&=({\mathcal T_{13}}+{\mathcal T_{23}})f(E)[1-f(E)].
\end{align}
\end{subequations}
As the transmission probability $\mathcal T_{ij}{\ge}0$ is positive, it is
straightforward to obtain the boundary of elastic transport coefficients as follows,
\begin{equation}
-1\le{K^{\rm el}_{12}/K^{\rm el}_{22}}\le 0,\quad
-1\le{K^{\rm el}_{12}/K^{\rm el}_{11}}\le 0.
\label{eq:el}
\end{equation}
The above expression is presented graphically by the red shadow regime in Fig.~\ref{fig:bound} .

While for a typical inelastic device consisting of three terminals, the Onsager coefficients are expressed as~\cite{Mahan}
\begin{subequations}
\begin{align}
&K^{\rm inel}_{11}=\left \langle E^2_1\right \rangle G^{\rm inel}_{11},\\
&K^{\rm inel}_{12}=\left \langle E_1\omega_3\right \rangle G^{\rm inel}_{12},\\
&K^{\rm inel}_{22}=\left \langle \omega^2_3\right \rangle G^{\rm inel}_{22}.
\end{align}
\end{subequations}
where the ensemble average over all inelastic processes is carried out as
\begin{equation}
\left \langle {\mathcal Q(E,\omega)} \right \rangle=\frac{\iint{dEd\omega{\mathcal Q(E,\omega)}G^{\rm inel}(E,\omega)}}{\iint{dEd{\omega}G^{\rm inel}(E,\omega)}},
\end{equation}
with $G^{\rm inel}=C_{\rm in}f_1(E_1)[1-f_2(E_2)]N_B(\omega_3)$. By applying the Cauchy-Schwarz inequality $\left \langle E^2_1\right \rangle\left \langle \omega^2_3\right \rangle-\left \langle E_1\omega_3\right \rangle^2\ge 0$, it is interesting to find that inelastic transport coefficients are bounded by
\begin{equation}
\frac{K^{\rm inel}_{11}}{K^{\rm inel}_{12}}{\times}\frac{K^{\rm inel}_{22}}{K^{\rm inel}_{12}}{\ge}1.
\label{eq:inel}
\end{equation}
We have provided a generic description of linear electronic and bosonic transport in the three-terminal geometry. Remarkably, the two simple relationships Eqs. \eqref{eq:el} and \eqref{eq:inel} hold for all thermodynamic systems in the linear-response regime.

\section{The basic model of inelastic thermoelectric transport: three-terminal double QD device}\label{DQD-device}

\begin{figure}[htb]
\begin{center}
\centering\includegraphics[width=7.5cm]{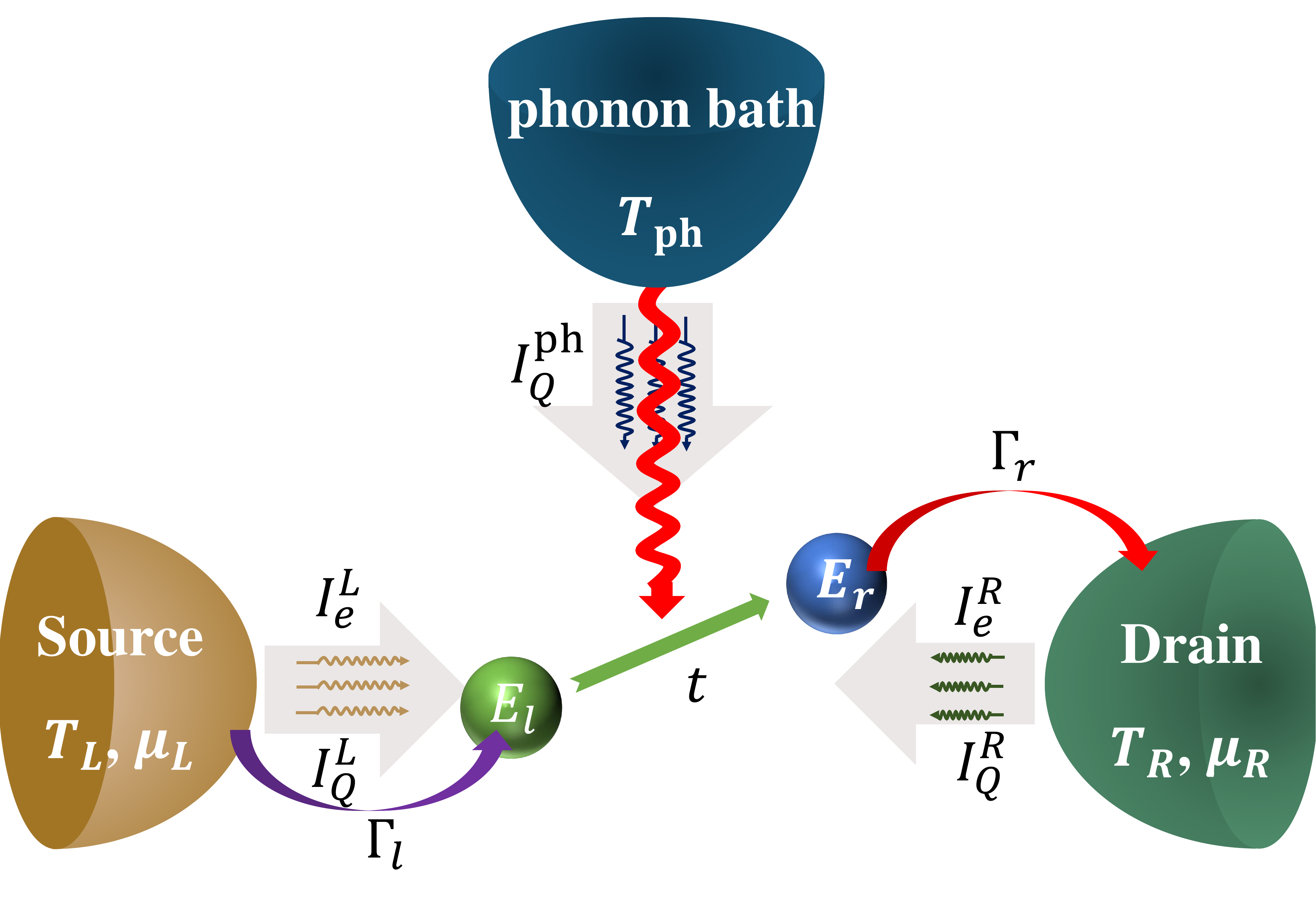}
\caption{Illustration of three-terminal inelastic transport. An electron left the source into the left QD (with energy $E_l$) hops to the right QD (with a different energy $E_r$) as assisted by a phonon from the phonon bath (with temperature $T_{\rm ph}$). The electron then tunnels into the drain electrode from the right QD. Such a process gives inelastic charge transfer from the source to the drain assisted by the phonon from the phonon bath. Both the process and its time-reversal contribute to the inelastic thermoelectricity in the system. The electrochemical potential and temperature of the source (drain) are $\mu_L$ and $T_L$ ($\mu_R$ and $T_R$), respectively.  $t$ is a hopping element between the QDs and $\Gamma_{l/r}$ are the hybridization energies of the dots to the source and drain electrodes, respectively.}~\label{fig:DQD}
\end{center}
\end{figure}

A typical inelastic thermoelectric device consists of three terminals: two electrodes (the source and the drain) and a boson bath (e.g., a phonon bath), which is schematically depicted in Fig.~\ref{fig:DQD}. In phonon-assisted hopping transport, the figure of merit is limited by the average frequency and bandwidth of the phonons (rather not electrons) involved in the inelastic transport~\cite{Jiangtransistors}.
Hartke {\it et al.}~\cite{HartkePRL} experimentally probes the electron-phonon interaction in a suspended InAs nanowire double QD,
which is electric-dipole coupled to a microwave cavity~\cite{2012Circuit,LiuPRL,GullansPRL,prete,dorschNL,Chen17,GuoPRApp21}.

Specifically, the system is described as the Hamiltonian
\begin{equation}
\hat H = \hat H_{\rm DQD} + \hat H_{\rm e-ph} + \hat H_{\rm lead} + \hat H_{\rm tun} + \hat H_{\rm ph},
\end{equation}
with
\begin{subequations}
\begin{align}
\hat H_{\rm DQD} &= \sum_{i=\ell,r} E_i \hat c_i^\dagger \hat c_i + ( t \hat c_l^\dagger \hat c_r + {\rm H.c.}) , \\
\hat H_{\rm e-ph} &= \gamma_{e-{\rm ph}} \hat c_l^\dagger \hat c_r (\hat a + \hat a^\dagger) + {\rm H.c.} ,\\
\hat H_{\rm ph} &= \omega_{0}\hat a^\dagger \hat a, \\
\hat H_{\rm lead} &= \sum_{j=L, R}\sum_{k} \varepsilon_{j,k} \hat c_{j,k}^\dagger \hat c_{j,k} ,\\
\hat H_{\rm tun} &= \sum_k V_{L, k} \hat c_\ell^\dagger \hat c_{L, k} + \sum_k V_{R, k} \hat c_r^\dagger \hat c_{R, k} +  {\rm H.c.},
\end{align}
\end{subequations}
where $\hat c_i^\dagger$ ($i=\ell, r$) creates an electron in the $i$-th QD with an energy $E_{i}$,
$ \gamma_{e-{\rm ph}}$ is the strength of electron-phonon interaction, and
$\hat a^\dagger$($\hat a$) creates (annihilates) one phonon with the frequency $\omega_{0}$.

For the three-terminal setup in Fig. ~\ref{fig:DQD},
nonequilibrium steady-state quantities of interest are the electric current $I_e$,  the electronic heat current traversing from the left reservoir to the right reservoir $I_Q^e=\frac{1}{2}(I_Q^L-I_Q^R)$, and the phonon heat current $I_Q^{\rm ph}$, with $I_Q^i$ ($i=L,R,{\rm ph}$) denoting the heat current flowing from the $i$th reservoir.
Specifically, the inelastic contribution to the currents is obtained from the Fermi golden rule of~\cite{Jiang2012},
\begin{equation}
I_e =eI_N, \,\,\,  I_Q^e=\frac{1}{2}(E_l+E_r)I_N, \,\,\, I_Q^{\rm ph} = (E_r-E_l) I_N.
\label{eq:currents}
\end{equation}
The current factor is $I_N = \Gamma_{l\rightarrow r} - \Gamma_{r\rightarrow l}$,
 and the transition rates are $\Gamma_{l\rightarrow r} \equiv \gamma_{e-{\rm ph}} f_{\ell} ( 1 - f_r) N_p^-$ and $\Gamma_{r\rightarrow l} \equiv \gamma_{e-{\rm ph}} f_r (1-f_{\ell}) N_p^+$,
  with $N_p^{\pm}=N_B+\frac{1}{2}\pm\frac{1}{2}{\rm sgn}(E_r-E_l)$ and the Bose-Einstein distribution for phonons $N_B\equiv [\exp({|E_r-E_l|}/{T_{\rm ph}})-1]^{-1}$. 

The thermodynamic affinities conjugated to those three currents satisfy the following relation~\cite{Onsager1,Onsager2}
\begin{equation}
{\dot S}_{\rm tot} = I_eA_1 + I_Q^eA_2 + I_Q^{\rm ph}A_3,
\end{equation}
where these conjugated affinities are
\begin{equation}
\begin{aligned}
A_1 &= \frac{\mu_L-\mu_R}{e}\left(\frac{1}{2T_L}+ \frac{1}{2T_R} \right), \, A_2=\frac{1}{T_R}-\frac{1}{T_L}, \\
A_3 &= \frac{1}{2T_L}-\frac{1}{2T_R}-\frac{1}{T_{\rm ph}}.
\end{aligned}
\end{equation}
Hence, based on the the phenomenological transport equations in the linear-response regime, the currents are reexpressed as~\cite{Jiang2012,JiangPRE,JiangJAP}
\begin{equation}
\begin{aligned}
\left( \begin{array}{cccc} I_e\\ I_Q^e \\ I_Q^{\rm ph} \end{array}\right) =
 \left( \begin{array}{cccc} G & L_1 & L_2 \\ L_1 & K_e^0 & L_3 \\ L_2 & L_3 & K_{pe}
        \end{array} \right) \left( \begin{array}{cccc} A_1 \\  A_2 \\ A_3
        \end{array}\right),
\label{eq:Onsager}
\end{aligned}
\end{equation}
And two thermopowers are defined as~\cite{Jiang2012}
\begin{equation}
S_1=\frac{L_1}{TG}, \quad S_2=\frac{L_2}{TG}.
\label{eq:thermopower}
\end{equation}

In the above Onsager matrix, $G$ denotes the charge conductivity, $L_1$ and $L_2$ represent the longitudinal and transverse thermoelectric effects~\cite{chen2021,zhouNM21}, respectively. $K_{e}^0$, $K_{pe}$ and $L_3$ are the diagonal and off-diagonal thermal conductance, which were originally derived based on the Onsager theory:
\begin{equation}
\begin{aligned}
& L_{1} = G\frac{E_l+E_r}{2e},  \quad   L_{2} = G\frac{E_r-E_l}{e}, \\
& K_{e}^0 = G\frac{(E_l+E_r)^2}{4e^2},  \quad    K_{pe} = G \frac{(E_r - E_r)^2}{e^2},  \\
& L_3 = G\frac{(E_l+E_r)(E_r-E_l)}{2e^2}.
\end{aligned}
\end{equation}
The conductance is $G=\frac{e^2}{k_BT}\Gamma_{1\rightarrow2}$, with $\Gamma_{1\rightarrow2}$ being the inelastic transition rate between the two QDs.  We assumed here that the coupling between the left QD and the source as well as that between the right QD the drain is much stronger than the coupling between the two QDs.

\section{Unconventional thermoelectric effects induced by inelastic transport}\label{inelastic-effects}
In this section, we show that how phonon-assisted inelastic transport leads to unconventional thermoelectric effects, such as the rectification effect, transistor effect, cooling by heating effect, and cooling by thermal current effect.

\subsection{Transistors and rectifiers}

Diodes and transistors are key components for modern electronics. In recent years, the manipulation and separation of thermal and electrical currents to process information in nano-scale devices have attracted tremendous interests~\cite{sdatta2005book,transistor0,Transistor1,YuPRE18,WangPRE,YuPRE19,transistor4,YuEntropy}. The design and experimental realization of the thermoelectric device present a striking first step in spin caloritronics, which concerns the coupling of heat, spin, and charge currents in magnetic thin films and other nanostructures~\cite{bauer2012spin}.  Meanwhile, phononic devices, which are devoted to the only use of heat currents for information processing, have also aroused extensive discussions over the past few decades~\cite{RenRMP}.

\begin{figure}[htb]
\begin{center}
\centering\includegraphics[width=8.0cm]{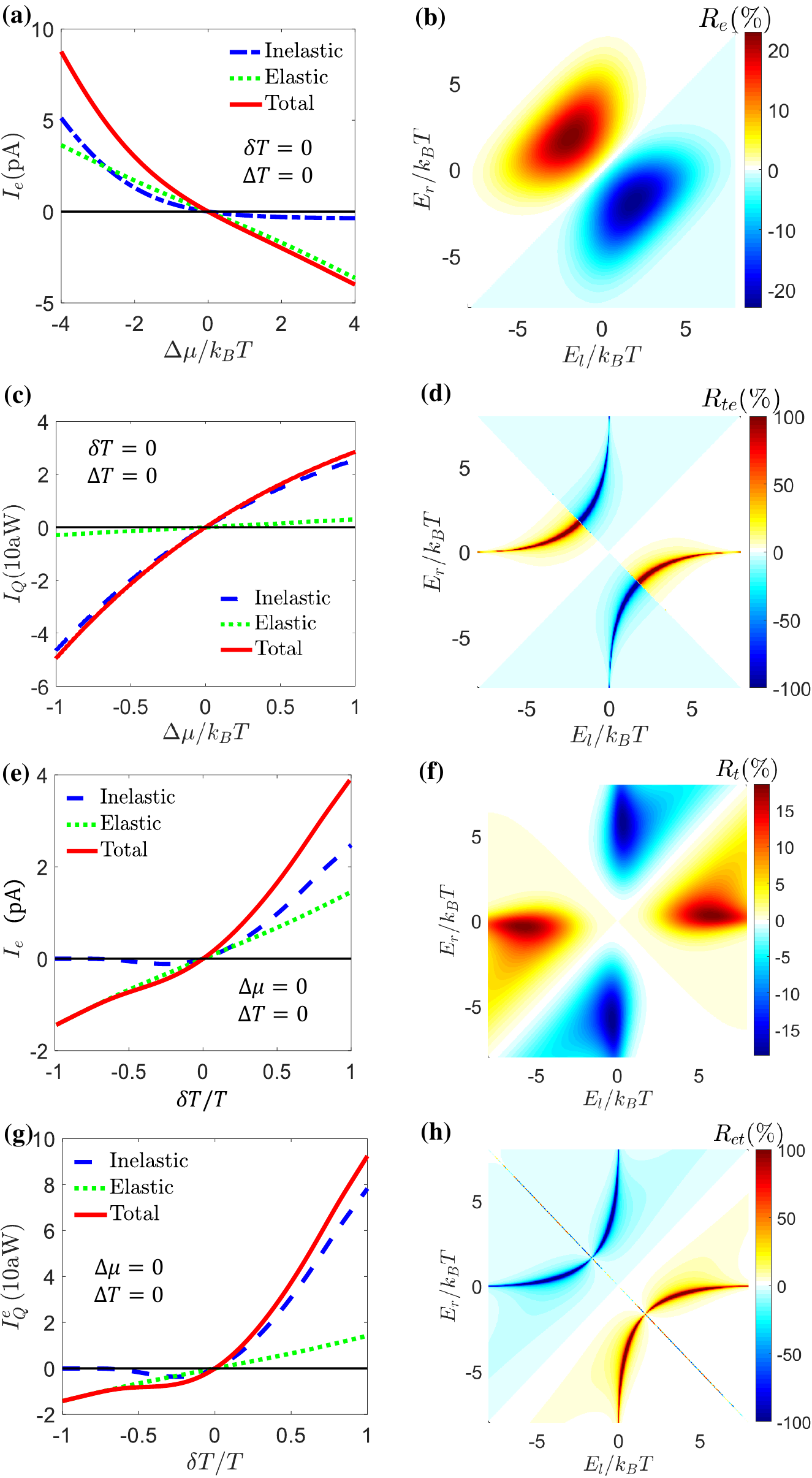}
\caption{Charge, heat,  and cross rectification effects. Figures are reproduced from Ref.~\cite{Jiangtransistors}.} \label{fig:rectification}
\end{center}
\end{figure}

In Ref.~\cite{Jiangtransistors}, we have shown that thermoelectric rectifier and transistor can be realized in the three-terminal double QD system, in which charge current and electronic and phononic heat currents are inelastically coupled.
Specifically, the coupled thermal and electrical transport allows standard rectification, i.e., charge rectification induced by a voltage bias. The magnitudes of the rectification effects are respectively
	defined by $R_e=\frac{I_e(V)+I_e(-V )}{|I_e(V )|+|I_e(-V )|}$ for
	charge rectification, $R_t=\frac{I_Q^e(\delta T)+I_Q^e(-\delta
		T)}{|I_Q^e(\delta T)|+|I_Q^e(-\delta T)|}$ for electronic heat
	rectification, $R_{et}=\frac{I_e(\delta T)+I_e(-\delta T)}{|I_e(\delta
		T)|+|I_e(-\delta T)|}$ for charge rectification induced by the temperature
	difference $\delta T$, and
	$R_{te}=\frac{I_Q^e(V)+I_Q^e(-V)}{|I_Q^e(V)|+|I_Q^e(-V)|}$ for heat
	rectification induced by voltage bias. The results displayed in Fig.~\ref{fig:rectification} including $I_e-V$, $I_Q^e-\delta T$, $I_Q^e-V$, and the $I_e-\delta T$ curves demonstrate significant rectification effects.

\begin{figure}[htb]
\begin{center}
\centering\includegraphics[width=8.0cm]{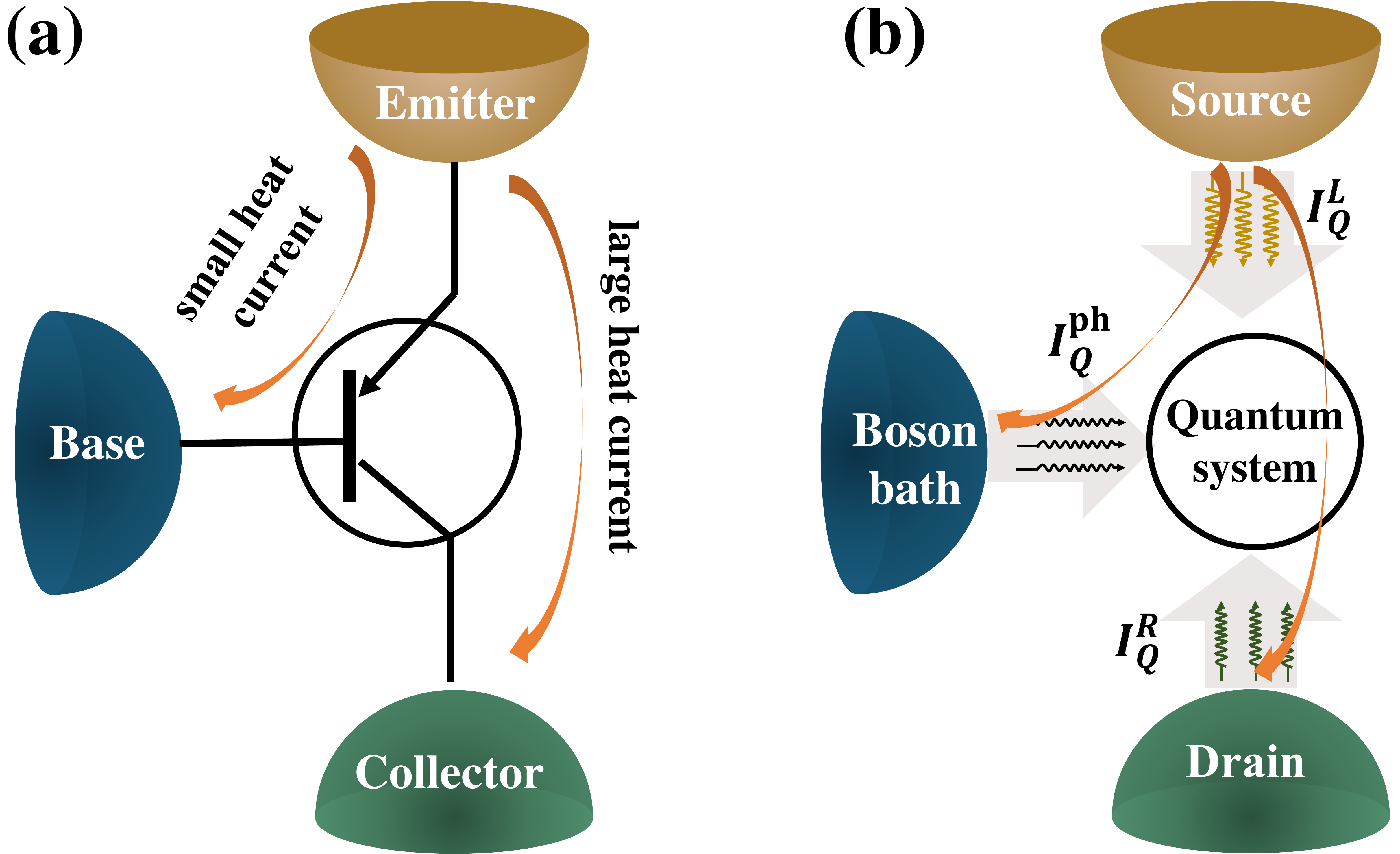}
\caption{(a) We describe the functioning of a conventional transistor here. A small current flows from emitter to base facilitating a large current from emitter to collector. The ratio of these currents, i.e, $\bar{\alpha}$ is the quantity that characterizes a transistor. (b) The specific three-terminal quantum-dot system that we propose as a transistor is represented here. The source lead, drain lead, and photon bath act like the emitter, collector, and base respectively. The small heat current flowing from the source to the boson bath, $I_Q^{\rm ph}$, can control the large heat current flowing from the source to the drain, $I_Q^{R}$. The ratio between the two heat currents defines the heat current amplification factor, $\bar{\alpha}$, which characterizes the thermal transistor effect. }~\label{fig:transistor}
\end{center}
\end{figure}

\begin{figure}[htb]
\begin{center}
\centering\includegraphics[width=8.0cm]{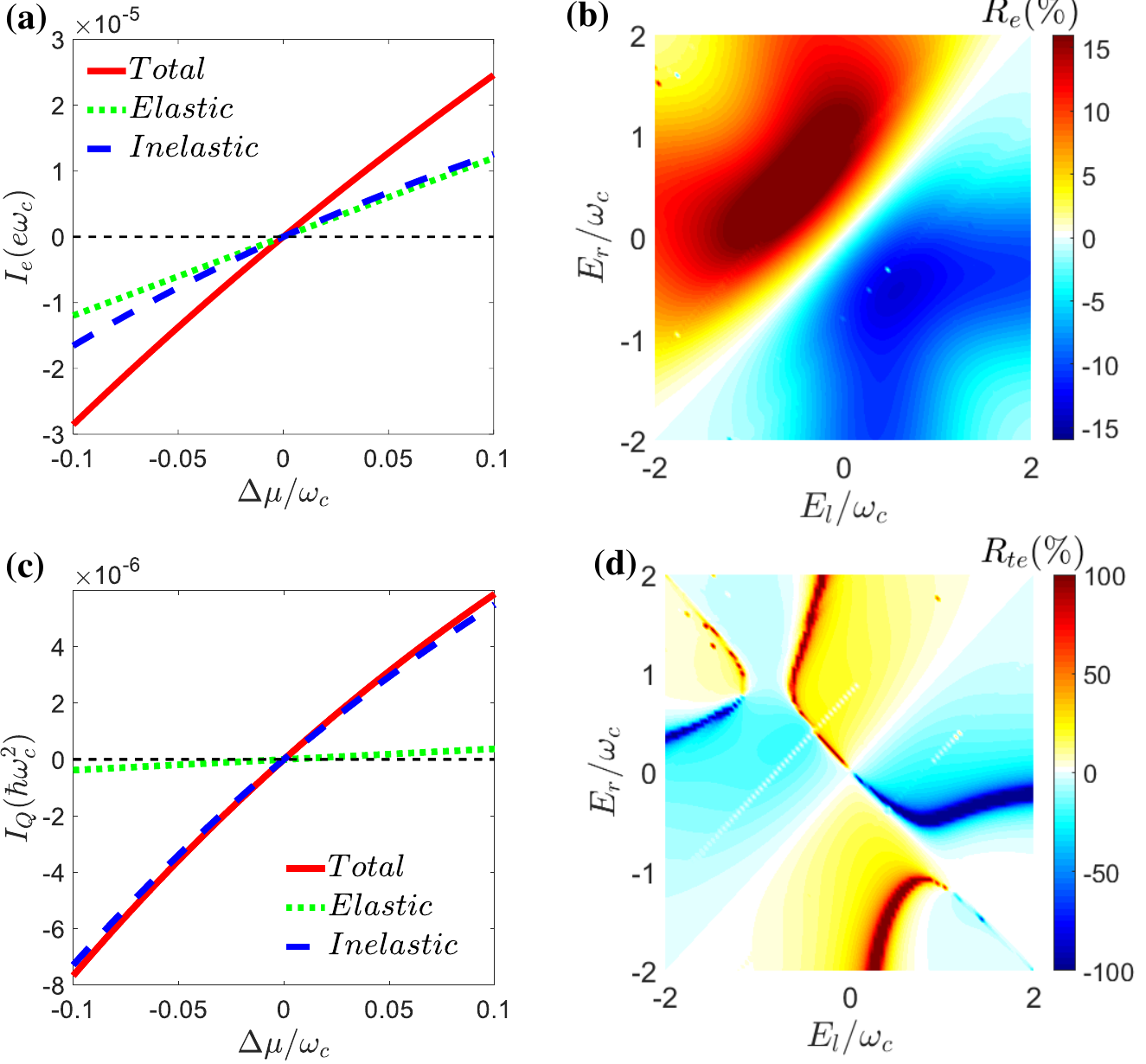}
\caption{The current (a) and the heat current (c) as the function of $\Delta \mu$. (b) Charge rectification $R_e$ and (d) cross rectification $R_{te}$ as the function of $E_l$ and $E_r$. Figures are reproduced from Ref.~\cite{MyPRBdiode}.}~\label{fig:rectification2}
\end{center}
\end{figure}

\begin{figure}[htb]
\begin{center}
\centering\includegraphics[width=9.0cm]{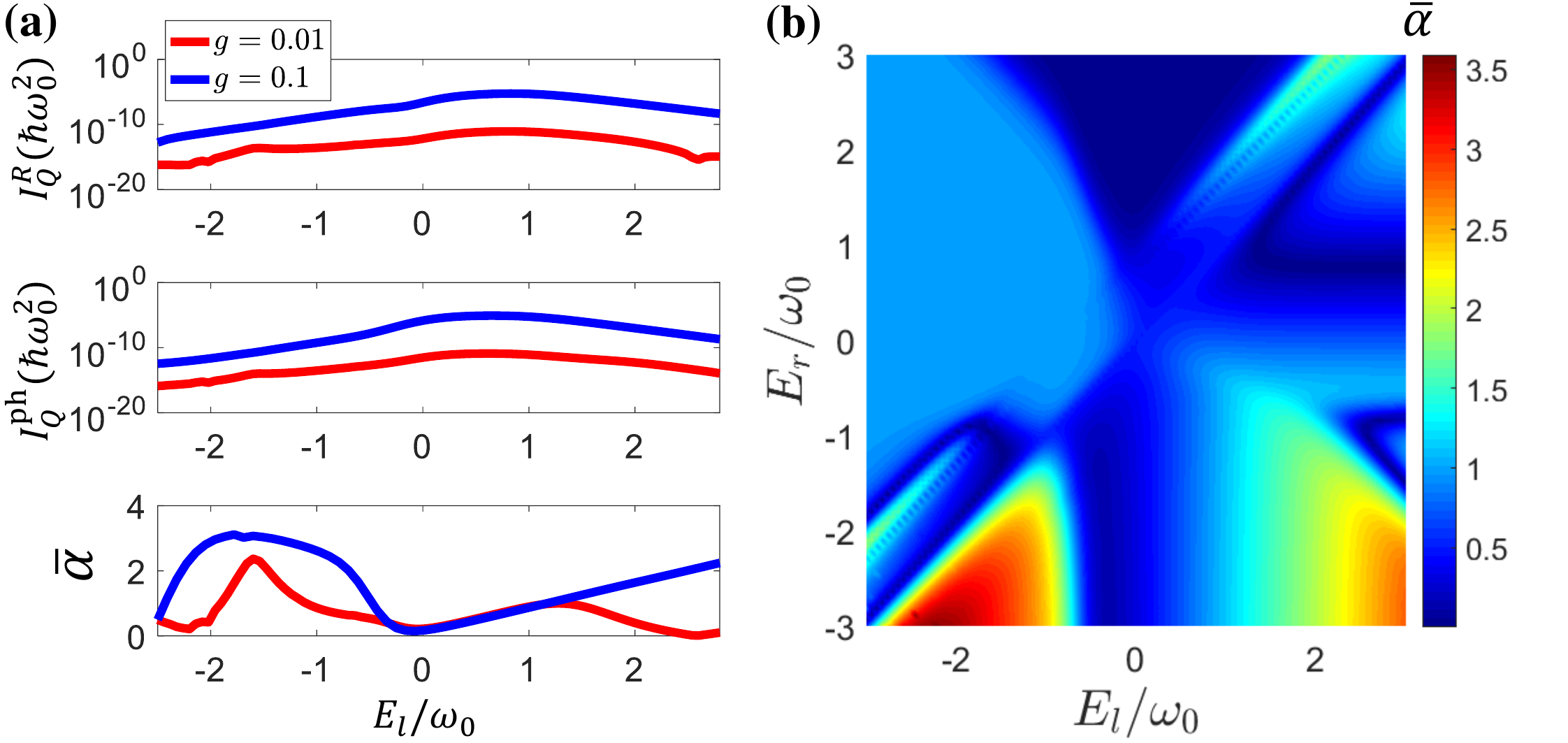}
\caption{(a) The heat current $I_Q^R$ flowing out from the right reservoir, photonic heat current $I_Q^{\rm ph}$ and heat current amplification factor $\bar{\alpha}$ as function of $E_l$ for different electron-photon interaction $g$. (b) The heat current amplification factor $\bar{\alpha}$ as the function of quantum dot energies $E_l$ and $E_r$. Figures are reproduced from Ref.~\cite{MyPRBdiode}. }~\label{fig:transistor-alpha}
\end{center}
\end{figure}

In addition to the diode effect, we further show that the three-terminal QD system is able to exhibit thermal transistor effect [see Fig. ~\ref{fig:transistor}]. It has been proposed that negative differential thermal conductance is compulsory for the thermal transistor effect~\cite{LiAPL06,RenRMP,WangPRE}. Here we remove such restriction on the thermal transistor effect,
 directly arising from the second law of thermodynamics.

From the phenomenological Onsager transport equation given by Eq. \eqref{eq:2OnsagerMatrix}, the average heat current amplification factor is then given by
\begin{equation}
\bar{\alpha}=\frac{\partial_{T_{\rm ph}}I_Q^L}{\partial_{T_{\rm ph}}I_Q^{\rm ph}}=\frac{K_{12}}{K_{22}}.
~\label{eq:alpha}
\end{equation}
It should be noted that $\overline\alpha$ only relies on the general expression of the transport coefficients $K_{12}$ and $K_{22}$. Specifically, for the elastic thermal transport, $\overline{\alpha}_{\rm el}$ is always below the unit as $-1<{K_{12}^{\rm el}}/{K_{22}^{\rm el}}<0$ (red shadow regime in Fig. ~\ref{fig:bound}). While for the inelastic case with the constraint coefficients bound at Eq. \eqref{eq:inel}, the average efficiency is given by $\overline{\alpha}_{\rm inel}<\left| {K_{11}^{\rm inel}}/{K_{12}^{\rm inel}}\right|$, which can be modulated in the regime $\overline{\alpha}_{\rm inel}$. Hence, the stochastic transistor may work as ${K_{11}^{\rm inel}}/{K_{12}^{\rm inel}}>1$.  Moreover, for the inelastic transport case, the Onsager coefficients are constraint by the second law of thermodynamics, $K_{11}K_{22}-K_{12}^2\ge0$. Therefore, the bound of amplification average efficiency is given by $0<\overline\alpha<\infty$ (blue shadow regime in Fig. ~\ref{fig:bound}).

A realistic example that achieves $\bar{\alpha}>1$ in the linear-response regime can be found in the three-terminal double QD system~\cite{Jiangtransistors}, which is expressed as
\begin{equation}
\bar{\alpha}=\left|\frac{E_l -\mu}{E_l - E_r}\right|.
\end{equation}
When $\left|E_l -\mu\right|>\left|E_l - E_r\right|$, $\bar{\alpha}$ can be greater than unity.
Therefore, we conclude that the thermal transistor effect then can also be realized in the linear response regime,
 in absence of the negative differential thermal conductance.

To further explain those phenomena, we expand the currents up to the second order in affinities
\begin{equation}
I_i=\sum_j M_{ij}A_j + \sum_{jk} L_{ijk}A_jA_k + {\mathcal O}(A^3),
\end{equation}
where $M_{ij}=M_{ij}^{\rm el} + M_{ij}^{\rm inel}$, with $M_{ij}$ denoting the linear-response coefficients and the second-order terms $L_{ijk}$ only shows up from the inelastic transport processes.
Practically, $M_{ij}$ and $L_{ijk}$ can be calculated with realistic material parameters~\cite{Jiangtransistors}. The first term on the right-hand side describes the linear response, whereas the second term gives the lowest-order nonlinear response.
The functionalities represented by various second-order coefficients are summarized in Table~\ref{table1}.

The influence of the strong electron-phonon interaction in thermoelectric transport is an intriguing research topic in nonequilibrium transport~\cite{PettaNRP,ZhuPRB03,JiangPRX,RenPRB12,Mi156,JinPRB21,WangPRR}.
However, the expression of currents in Eq.~(~\ref{eq:currents}) may break down as the electron-phonon interaction becomes strong, where the high-order electron-phonon scattering processes should be necessarily included to properly characterize the electron current and energy current.
Alternatively, the strong light-matter interaction also provides an excellent way for designing efficient thermoelectric devices. In Ref.~\cite{MyPRBdiode}, we show that significant  rectification effects (including charge and Peltier rectification effects) [see Fig.~\ref{fig:rectification2}] and linear thermal transistor effects [see Fig.~\ref{fig:transistor-alpha}] can be enhanced due to the nonlinearity induced by the large electron-photon interaction in circuit-quantum-electrodynamics systems. The above results show that the synergism of electronics and boson in open systems can provide a novel solution for seeking high-performance thermoelectric devices and information storage technology in the future.

\begin{table}[htb]
\caption{Functionality of second-order coefficients}
\begin{tabular}{llllllllll}\hline 
Terms ($L_{ijk}$) & \mbox{} & Diode or Transistor effect \\
\hline
$L_{111}$ &  \mbox{} & charge rectification \\
$L_{222}$, $L_{333}$ &  \mbox{} & electronic and phononic heat rectification \\
$L_{233}$, $L_{322}$ &  \mbox{} & off-diagonal heat rectification \\ 
$L_{122}$, $L_{133}$ &  \mbox{} & charge rectification by temperature difference\\
$L_{211}$, $L_{311}$ &  \mbox{} & heat rectification by voltage bias\\
$L_{113}$, $L_{123}$ &  \mbox{} & boson-thermoelectric transistor \\
$L_{212}$, $L_{112}$ &  \mbox{} & other nonlinear thermoelectric effects\\
\hline 
\end{tabular}
\label{table1}
\end{table}

\begin{figure}[htb]
\begin{center}
\centering\includegraphics[width=8.0cm]{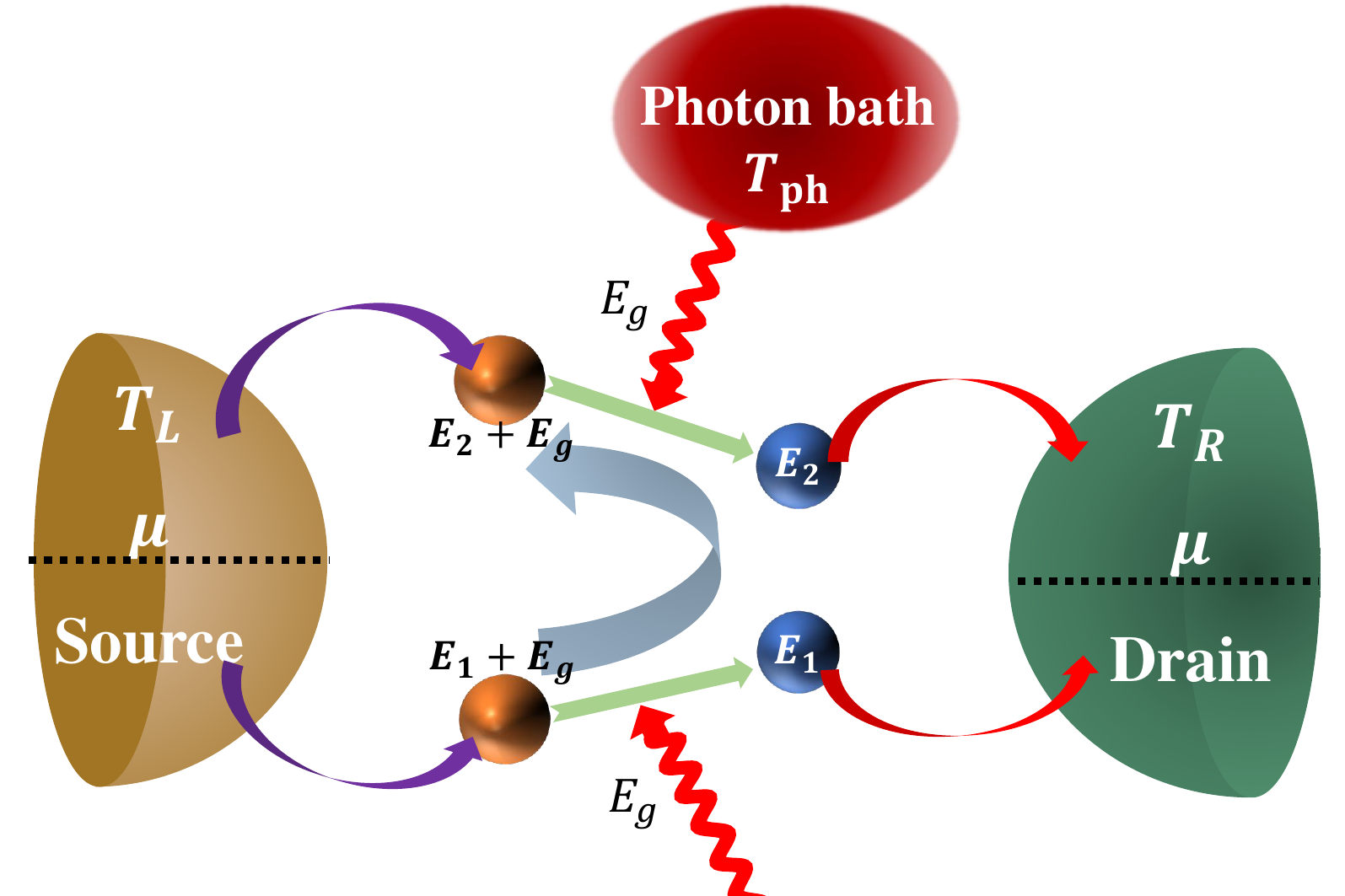}
\caption{Schematic of the ``cooling by heating'' refrigerator. The source and drain (with different temperatures $T_{L(R)}$, and electrochemical potentials $\mu_L=\mu_R\equiv\mu$), are connected by two quantum dots, each having two discrete energy levels. }~\label{fig:cooling-heating}
\end{center}
\end{figure}

\subsection{Cooling by heating effects}

According to Clausius' second law of thermodynamics, we know that heat cannot spontaneously transfer from the cold reservoir to the hot reservoir~\cite{Maxwellbook}. Usually, the second law is expressed in a two-terminal system. For three-terminal systems, the second law of thermodynamics has a more complex case where some counterintuitive effects can be allowed~\cite{FabioPRB18,TaddeiPRB1,SegalPRE19,JordanPRB20,SegalPRE21}. For example, in Ref.~\cite{Cooling2}, Cleuren {\it et al.} proposed that one cold reservoir can be cooled by two hot reservoirs without changing the rest of the world due to the transport mechanism of inelastic scattering, which is termed as ``cooling by heating'' effect.

As exemplified in Fig.~\ref{fig:cooling-heating}, to perform cooling by heating a device must have three reservoirs (source, drain, and photon bath) and two adjoining quantum dots. Each quantum dot has a lower and upper energy level. The source is kept at ambient temperature $T_{L}$, photon bath is hotter with $T_{\rm ph}>T_L$,
and drain is colder with $T_R<T_L$.
The device then utilizes the heat flowing from the photon bath to the source to ``drag" heat out of the drain,
even though the drain is colder than the other two hot reservoirs.

The basic mechanism is that under the influence of high-temperature photons, the electrons with energy lower than the Fermi level in the source will inelastically pass through the lower energy regime of the two quantum dots and tunnel into the drain.
Similarly, the electrons with energy higher than the Fermi level in the drain will transport into the source through the higher energy regime of the two quantum dots. Simultaneously, the electron needs to absorb one photon to complete the cyclic transition process between two quantum dots. The cooling by heating effect in quantum systems can be understood that as the quantum device is driven by the external work, the heat is extracted from the cooling reservoir and absorbed by the hot reservoir.

The efficiency of cooling by heating device (refrigerator) is defined as the heat current flowing out of the drain (the drain being refrigerated) divided by the heat current flowing out of the photon bath, i.e., $\eta_{\rm CBH}={I_Q^R}/{I_Q^{\rm ph}}$. The upper bound on such a refrigerator efficiency is given by the condition that no entropy is generated.
Then, the corresponding efficiency is given by
\begin{equation}
\eta^{\rm rev}_{\rm CBH} = \frac{T_R(T_{\rm ph} - T_L)}{T_{\rm ph}(T_L-T_R)}.
\end{equation}
Meanwhile, the refrigerator reaches the reversible regime with both heat currents $I_Q^R$ and $I_Q^{\rm ph}$ vanishing simultaneously, while it contains a nonzero cooling efficiency
\begin{equation}
\eta_{\rm CBH} = \frac{E_2-E_1}{2E_g},
\end{equation}
with the $E_1$ and $E_2$ being the energies of the right quantum dot, and $E_g$ being the energy gap between the upper (down) levels. It is worth noting that the increase of the  entropy rate of the whole system is not negative,
and the system satisfies the second law of thermodynamics for the entropy reduction in the source and drain is compensated by the larger entropy increase in the photon bath.

\subsection{Cooling by heat current effects}

\begin{figure}[htb]
\begin{center}
\centering\includegraphics[width=8.0cm]{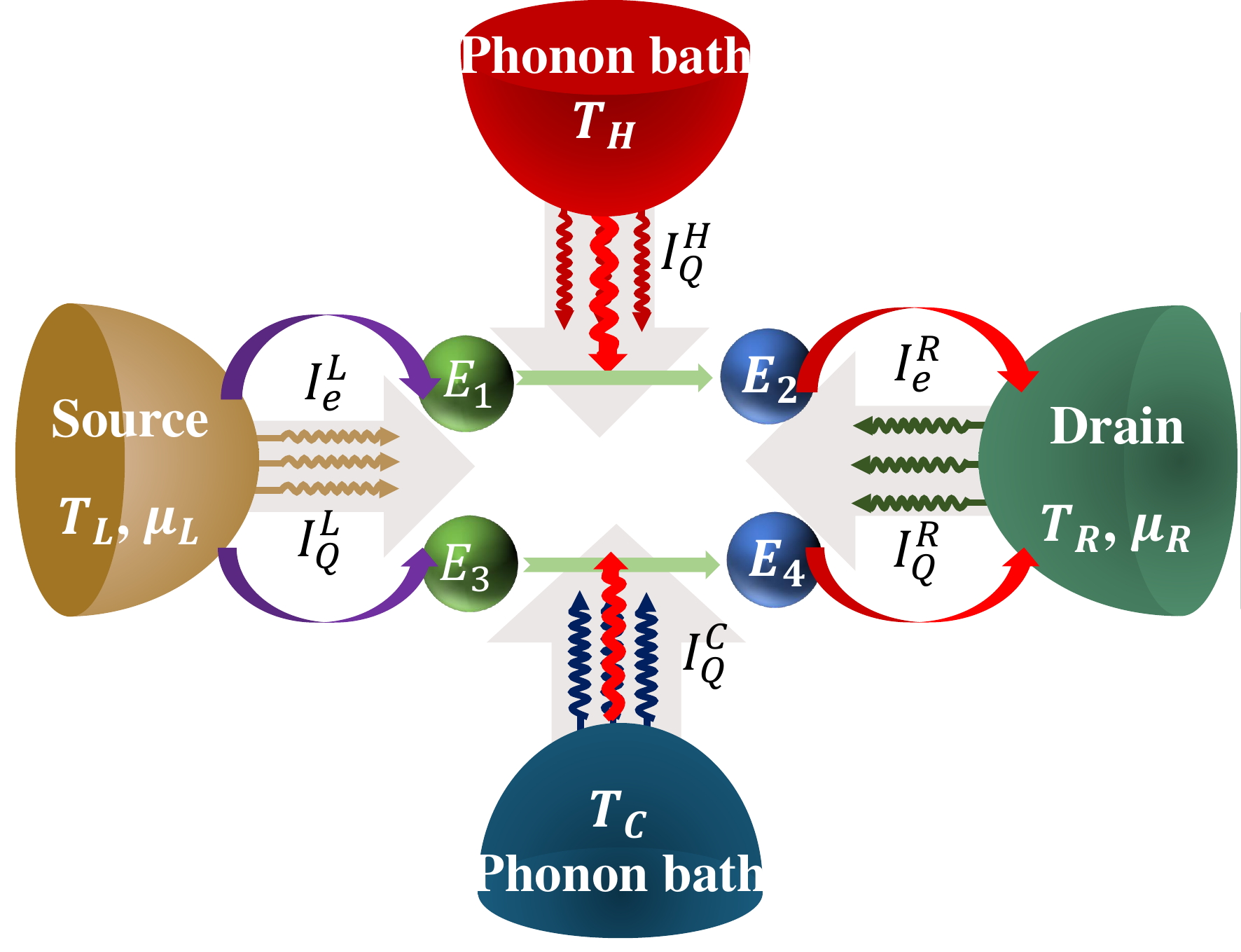}
\caption{Schematic of quantum dots four-terminal thermoelectric devices. There are two parallel transport channels. Each channel has two quantum dots with different energies and a heat bath to enable inelastic transport. The two channels are spatially separated so that the heat bath $H$ ($C$) couples only to the upper (lower) channel. Four heat currents $I_Q^L$, $I_Q^R$, $I_Q^C$, $I_Q^H$ and the electric currents $I_e$ are illustrated. }~\label{fig:4T-terminal}
\end{center}
\end{figure}

In this section, we show that a nontrivial phonon drag effect, termed by ``cooling by heat current''~\cite{MyPRBdemon}, can emerge in four-terminal QD thermoelectric systems with two electrodes and two phonon baths, shown in Fig. ~\ref{fig:4T-terminal}.
The source (or the drain) can be cooled by passing a thermal current between the two phonon baths, without net heat exchange between the heat baths and the electrodes. This effect, which originates from the inelastic-scattering process, could improve the cooling efficiency and output power due to spatial separation of charge and heat transport~\cite{Ora2015,MazzaPRB}.

\begin{figure}[htb]
\begin{center}
\centering \includegraphics[width=8.5cm]{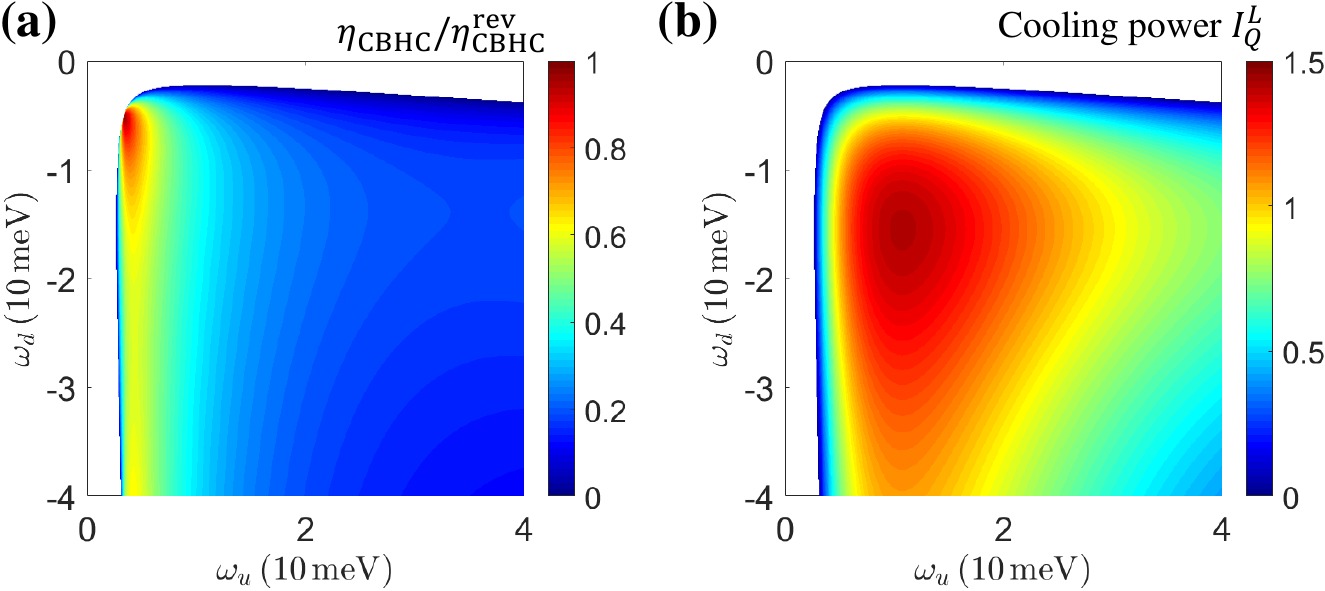}
\caption{(a) COP and (b) cooling power of the cooling by transverse heat current effect as functions the two energies $\omega_{u}$ and $\omega_{d}$. The white areas represent the parameter regions where the cooling by heat current effect cannot be achieved, i.e., $I_Q^L<0$. Figures are reproduced from Ref.~\cite{MyPRBdemon}}~\label{fig:PE-ud}
\end{center}
\end{figure}

Specifically, the system consists of four quantum dots: QDs 1 and 2 with electronic energy $E_1$ and $E_2$ are coupled with the hot heat bath $H$, while QDs 3 and 4 with energy $E_3$ and $E_4$ are coupled with the cold heat bath C.
there are one electrical current $I_e$ flowing from the source to drain and four heat currents, $I_Q^L$, $I_Q^R$, $I_Q^H$, and $I_Q^C$. Due to energy conservation~\cite{Onsager1,Onsager2}, i.e., $I_Q^L + \frac{\mu_L}{e}I_e + I_Q^H + I_Q^C + I_Q^R - \frac{\mu_R}{e}I_e=0$, the entropy production of the whole system is given by~\cite{JiangPRE}
\begin{equation}
\frac{dS}{dt} = I_Q^L A_S + I_Q^{\rm in}A_{\rm in} + I_Q^qA_q + I_eA_e,
\end{equation}
and the affinities are defined as
\begin{equation}
\begin{aligned}
&A_S \equiv \frac{1}{T_R} - \frac{1}{T_L}, \quad A_{\rm in} \equiv \frac{1}{T_R} - \frac{1}{2T_H} - \frac{1}{2T_C}, \\
&A_q \equiv \frac{1}{T_C} - \frac{1}{T_H}, \quad A_e \equiv \frac{\mu_L-\mu_D}{eT_R}.
\end{aligned}
\end{equation}
$I_Q^{\rm in}= I_Q^H+I_Q^C$ is regarded as the total heat current injected into the central quantum system from the two thermal baths. $I_Q^q=(I_Q^H-I_Q^C)/2$ is the exchanged heat current between the two heat baths intermediated by the central quantum system.
$T_i$ ($i=L,R,H,C$) are the temperatures of the four reservoirs, respectively.
Here, we restrict our discussions on situations where there is only one energy level in each QD that is relevant for the transport.

In this regime, the heat currents derived from the Fermi golden rule~\cite{Jiangtransistors} can be written as,
\begin{equation}
\begin{aligned}
I_Q^L &= E_1I_{12} + E_3I_{34}, \quad I_Q^H = \omega_uI_{12}, \quad I_Q^C = \omega_dI_{34},\\
I_Q^{\rm in} &= \omega_uI_{12} + \omega_dI_{34}, \quad I_Q^q = \frac{1}{2}(\omega_uI_{12} - \omega_dI_{34}).
\end{aligned}
\end{equation}
Here $I_{12}= \Gamma_{1\rightarrow 2} - \Gamma_{2\rightarrow 1}$ ($I_{34}= \Gamma_{3\rightarrow 4} - \Gamma_{4\rightarrow 3}$) is the phonon-assisted hopping particle currents through the up (down) channel. $\Gamma_{i\rightarrow j}$ is the electron transfer rate from QD $i$ to QD $j$~\cite{Jiang2012,Jiang2013}, and $\omega_u = E_2-E_1$, $\omega_d= E_4-E_3$ denoting the QDs energy difference in the up and down channel.

We note that the source by driving a heat current between the heat baths $H$ and $C$, i.e., ``cooling by heat current effect'', is different from the above ``cooling by heating effect'' where cooling is driven by a finite heat current injected into the quantum system. In the cooling by heat current effect, heat injected into the quantum system is not necessary, since the driving force of the cooling is the energy exchange between the two heat baths via the central quantum system.

For convenience, we demonstrate the cooling by heat current effect in the situations with $A_e=A_{\rm in}=0$. The coefficient of performance (COP) in our four-terminal system can be given by~\cite{trade-off,MyPRBdemon}
\begin{equation}
\eta_{\rm CBHC}=\frac{I_Q^L}{I_Q^q}.
\end{equation}
The reversible COP is $\eta^{\rm rev}_{\rm CBHC} =-{A_q}/{A_S}$. We show how the cooling power $I_Q^L$ and COP $\eta_{\rm CBHC}$ vary with the two energies, $\omega_{u}$ and $\omega_{d}$ in Fig.~\ref{fig:PE-ud}. Both the COP $\eta_{\rm CBHC}$ and the cooling power $I_Q^L$ favor the situations with $-\omega_{u}>\omega_{d}$. For such a regime, cooling induced by the cold terminal $C$ is more effective, for each phonon emission process provides more energy to the heat bath $C$.

\begin{figure}[htb]
\begin{center}
\centering \includegraphics[width=8.0cm]{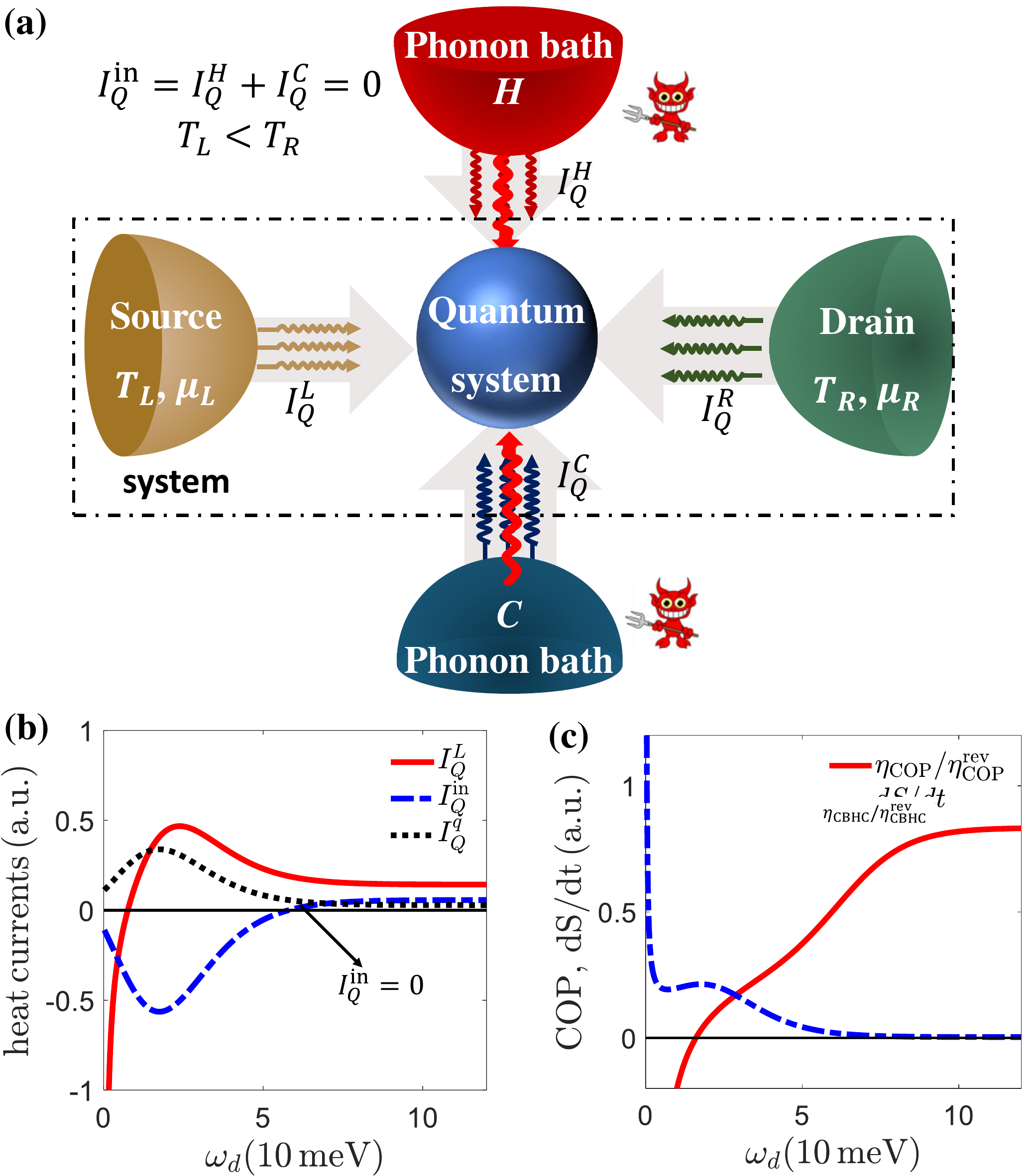}
\caption{(a) Schematic of a four-terminal mesoscopic thermoelectric device as a Maxwell demon. The demon supplies no work or heat to the system, i.e., the total heat current injected into the central quantum system from the two thermal baths is zero, $I_Q^{\rm in}=I_Q^H+I_Q^C=0$. (b) Thermal currents, and (c) coefficient of performance ratio $\eta_{\rm CBHC}/\eta^{\rm rev}_{\rm CBHC}$ and entropy production $dS/dt$ as a function of the QD energy $\omega_{d}$. Only when the case with $I_Q^{\rm in}=0$ represents that the nonequilibrium Maxwell demon. Figures are reproduced from Ref.~\cite{MyPRBdemon}.}~\label{figure:demon}
\end{center}
\end{figure}

As shown in Fig.~\ref{figure:demon}, we further find that the cooling by heat current effect can indeed exist when the total heat current injected into the quantum system vanishes (i.e., $I_Q^{\rm in}=0$), which termed as ``Maxwell demon''~\cite{SanchezPRRes,SanchezDemon,demonDQD,MyPRBdemon,demon1,demon2,demon3,demon4,MyCPL21}. The Maxwell demon based on two nonequilibrium baths (the cold and hot baths) can reduce the entropy of the system (the source and the drain), without giving energy or changing the particle number of the system. More specifically, the heat current can flow from the cold bath to the hot one without external energies or changing the number of particles in the system.

\section{Enhancing three-terminal thermoelectric performance using nonlinear transport effects}

Nonlinear transport effects can enhance elastic and inelastic thermoelectric efficiency and power when the voltage and/or temperature bias is large~\cite{Jiang2017}.
The reason is that linear-response theory usually fails when the voltage and/or temperature bias on the scale of the electrons' relaxation length (typically given by the electron-electron or electron-phonon scattering length) is comparable to the average temperature.
This point is particularly important for many thermoelectric applications.
{In particular,  S\'anchez  {\it et al.} based on the seminal works~\cite{SanchezCRP,David2011PRB,DavidPRL,SanchezPRB13,David-refrigerator},
investigated nonlinear quantum transport through nanostructures and mesoscopic systems driven by thermal gradients or in combination with voltage biases.
Specifically, when the temperature of the phonon bath increases,
the nonlinear thermoelectric transport leads to significant improvement of both the heat-to-work energy efficiency and the output electric power. All these effects are found to be associated with inelastic and elastic thermoelectric contributions.}

\subsection{Effects of nonlinear transport on efficiency and power for elastic thermoelectric devices}

We study the nonlinear transport effects on the performance of elastic thermoelectric devices. A simple candidate of such devices is a two-terminal QD thermoelectric device, i.e., a QD  with energy $E_0$ connected with the source (of temperature $T_h$) and the drain (of temperature $T_c<T_h$) electrodes via resonant tunneling~\cite{BijayPRL21,JunjiePRL}.
The electrical and heat currents can be calculated using the Landauer formula~\cite{Sivan,butcher1990,BENENTI20171}
\begin{subequations}
\begin{align}
& I_e = e\int \frac{dE}{2\pi} {\cal T}_e(E) [f_L(E) - f_R(E)] , \\
& I_{Q}^e = \int \frac{dE}{2\pi} (E-\mu_L) {\cal T}_e(E) [f_L(E) - f_R(E)],
\end{align}
\end{subequations}
with the energy-dependent transmission function ${\cal T}_e(E)=\frac{\gamma_e^2}{(E-E_0)^2 + \gamma_e^2}$.

Here, we consider harvesting the heat from the hot reservoir to generate electricity. The energy efficiency is hence described as
\be
\eta_{\rm HE} = \frac{P_{\rm HE}}{Q_{\rm in}} \le \eta_C,
\ee
with $T_h=T_{\rm ph}$ and $T_c=T_L=T_R$. The output power is
\be
P_{\rm HE}=-I_e V.
\ee
with $\mu_L=eV/2=-\mu_R$. The heat injected into the system from the hot reservoir is given by
\be
Q_{\rm in} = I_{Q}^e + I_{Q}^{\rm pr},
\ee
with $I_{Q}^{\rm pr}$ being the parasitic phonon heat current~\cite{Jiang2017}.

\subsection{Nonlinear transport enhances efficiency and power for inelastic thermoelectric devices}

We study the energy efficiency and output power of a double-QDs three-terminal thermoelectric device in the nonlinear transport regime. The device is schematically depicted in Fig. ~\ref{fig:DQD}. Here, we consider harvesting the heat from the phonon bath to generate electricity.
The heat injected into the system from the photon bath is given by
\be
Q_{\rm in} = I_{Q}^{\rm ph} + I_{Q}^{\rm pr} ,
\ee
where $I_{Q}^{\rm ph}= 2(E_r-E_l)\Gamma_{12}$ is the phononic current flowing from the phonon bath and
$\Gamma_{12}$ is the rate of electron transfer from the left QD to the right QD due to the electron-phonon scattering.
 The electrical current is given by
\begin{equation}
I_e = 2 e \gamma_e [ f_L(E_l) - f_1] + 2 e \gamma_e^\prime [f_L(E_r) - f_2] ,
\label{fullC}
\end{equation}
where the factor of two in the above equation comes from electron spin degeneracy.
$f_i$ ($i=1,2$) are the probabilities of finding an electron on the $i$th QD,
and they are determined by the nonequilibrium steady-state distributions on the QDs,
\begin{equation}
\begin{aligned}
0 = \frac{d f_1}{dt} =  - \gamma_e [ f_1 - f_L(E_l)] - \gamma_e^\prime [ f_1 - f_R(E_l)] - \Gamma_{12} , \\
0 = \frac{d f_2}{dt} =  - \gamma_e [ f_2 - f_R(E_r)] - \gamma_e^\prime [ f_2 - f_L(E_r)] + \Gamma_{12} .
\end{aligned}
\end{equation}
$\gamma_e/\gamma_e^\prime$ is the tunneling between the QD and the reservoir.
The linear transport coefficients are obtained by calculating the ratios between currents and affinities in the regime with very small voltage bias and
temperature difference [see Eq.~\eqref{eq:Onsager}].

\begin{figure}[htb]
\begin{center}
\centering\includegraphics[width=8.5cm]{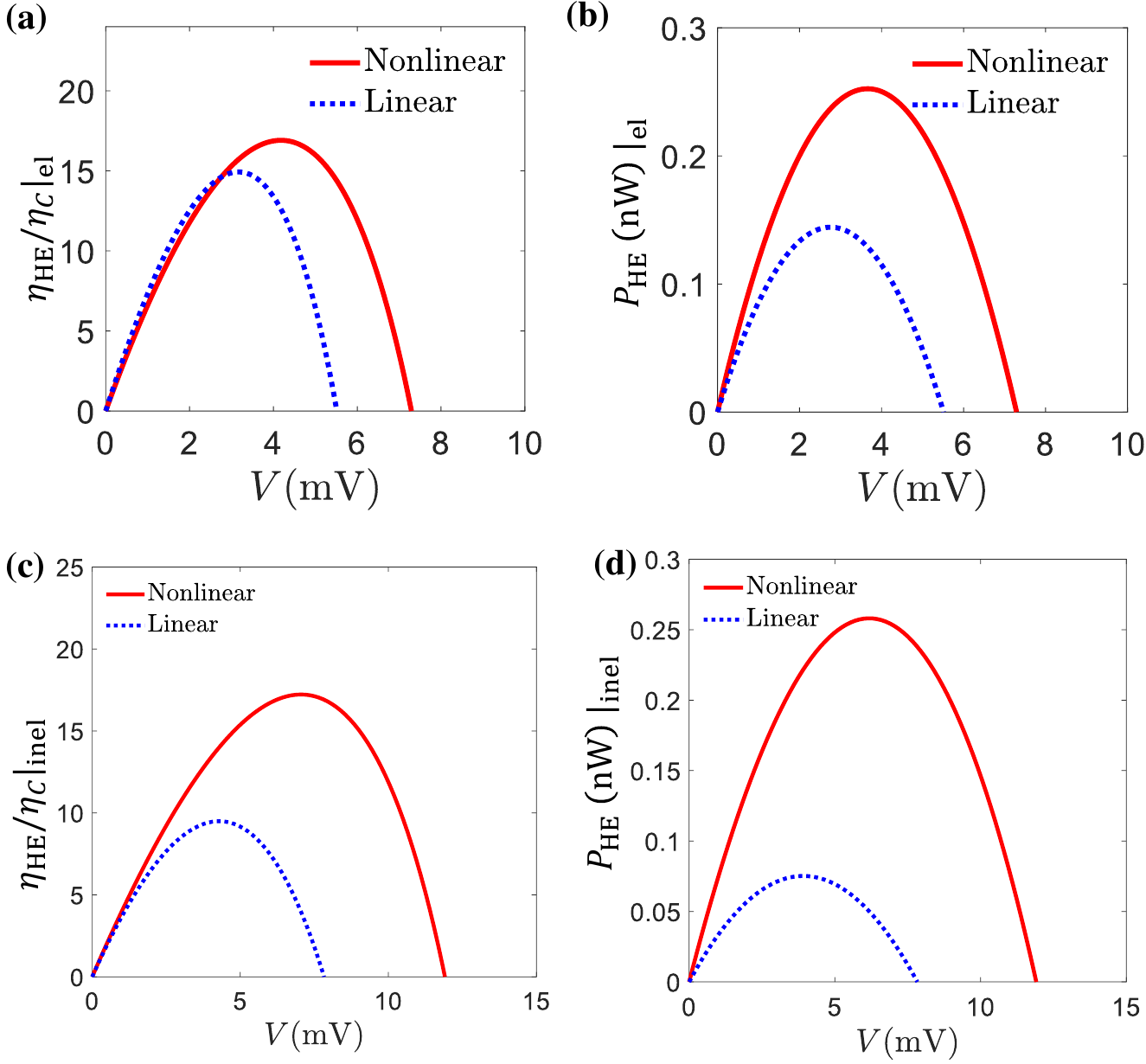}
\caption{(a)-(b). Elastic thermoelectricity. (a) Energy efficiency $\eta_{\rm HE}/\eta_C$ and (b) output power $P_{\rm HE}$ as a function of voltage $V$ (in units of mV) for nonlinear and linear transport. (c)-(d) Inelastic thermoelectricity. (a) Energy efficiency $\eta_{\rm HE}/\eta_C$ and (b) output power $P_{\rm HE}$ as functions of voltage $V$ (in units of mV) for the inelastic thermoelectric device. Figures are reproduced from Ref.~\cite{Jiang2017}.}
\label{Nonlinear-Figure1}
\end{center}
\end{figure}

In Fig.~\ref{Nonlinear-Figure1}, we perform a comparative study of the nonlinear transport effect on the maximum efficiency and power for inelastic and elastic thermoelectric devices systematically.
We find that the nonlinear effect can significantly improve the performance of thermoelectric devices, e.g., thermodynamic efficiency and output power, both for elastic and inelastic cases.

\section{Enhancing efficiency and power of three-terminal device by thermoelectric cooperative effects}

\begin{figure}[htb]
\includegraphics[height=5.0cm]{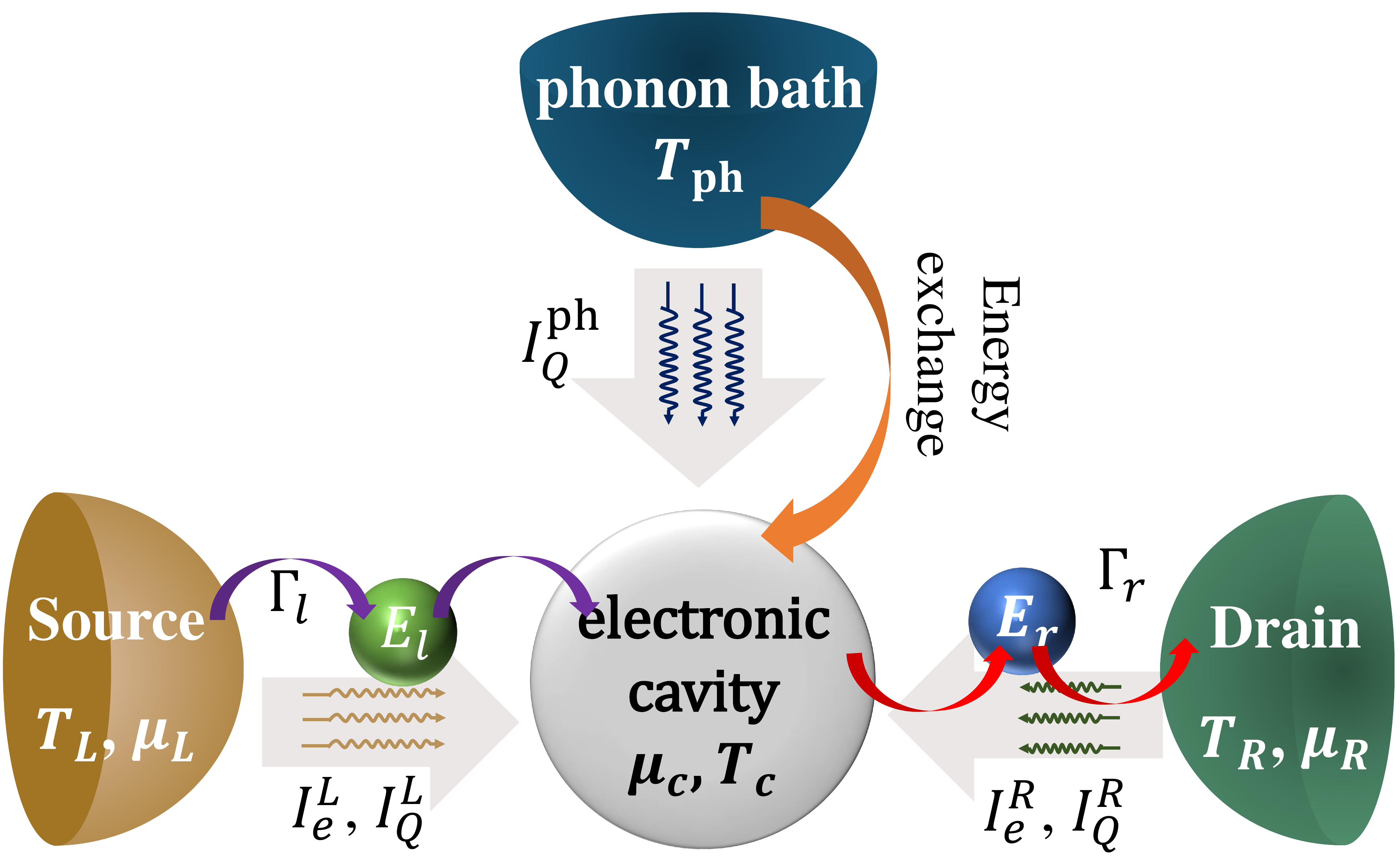}
\caption{Schematic view of a three-terminal thermoelectric system. The three-terminal device is composed by two electronic reservoirs and a phonon bath, which is held at $T_{\rm ph}$ temperature, The central cavity, which is thermalized by the phonon bath, is connected to two electrodes via two quantum dots at energy $E_{l/(r)}$.}
\label{JAP-model}
\end{figure}

In the following section, we discuss how the efficiency and output power of the three-terminal heat device can be enhanced by the thermoelectric cooperative effect in the linear-response regime. We consider the setup shown schematically in Fig. ~\ref{JAP-model}, which consists of two electronic reservoirs and a phonon bath. The central cavity, which is warmed up by the phonon bath, is connected to two electrodes via two QDs at energy $E_{l(r)}$. There are two thermoelectric effects, one of which belongs to inelastic processes, while the other exists in the elastic process. These two effects are related to two temperature gradients and correspond to the transverse and longitudinal thermoelectric effects, respectively. We show that the energy cooperation between the transverse and longitudinal thermoelectric effects in the three-terminal thermoelectric systems can lead to markedly improved performance of the heat device.

A full description of the thermoelectric transport in three-terminal systems is given by Eq. \eqref{eq:Onsager}. The cooperative effects in the thermoelectric engine can be elucidated by a geometric interpretation~\cite{JiangJAP,MyJAP,CPB}. The two temperature differences can be parametrized as
\begin{equation}
\delta T = T_A \cos\theta, \quad \Delta T = T_A \sin\theta .
\end{equation}
At given $\theta$, the figure of merit is given by
\begin{equation}
ZT = \frac{G S_{\rm eff}^2 T^2}{K_{\rm eff} - GS_{\rm eff}^2 T^2}.
\label{zzt}
\end{equation}
Here, $S_{\rm eff}=S_1\cos\theta + S_2\sin\theta$ and $K_{\rm eff}=K_e^0\cos^2\theta+2L_3\sin\theta\cos\theta+K_{pe}\sin^2\theta$.
$S_1$ and $S_2$ given by Eq. \eqref{eq:thermopower} denote the longitudinal and transverse thermopowers, respectively.
Then, the ``second-law efficiency'' of the thermoelectric engine is expressed as
\begin{equation}
\begin{aligned}
\phi = \frac{-I_e^LV}{I_{Q}^L A_2 + I_{Q}^{\rm ph}A_3} \le \phi_{\max} = \frac{\sqrt{ZT+1}-1}{\sqrt{ZT+1}+1},
\end{aligned}
\end{equation}
which is defined by the output free energy divided by the input free energy~\cite{JiangPRE,HajilooPRB,ManzanoPRR}. The rate of variation of free energy associated with a current is given by the product of the current and its conjugated thermodynamic force.
Hence, the denominator of the above equation consists of heat currents multiplied by temperature differences.
Such free-energy efficiencies have been discussed for near-equilibrium thermodynamics (in the linear response regime) or arbitrarily far from equilibrium, ranging from biological~\cite{Caplan} to quantum Hall system~\cite{SanchezPRL,SanchezDemon,ArracheaPRL,HajilooPRB}.

Upon optimizing the output power of the thermoelectric engine, one obtains $W_{\max} = \frac{1}{4} P_F T_A^2$, with the power factor
\begin{equation}
P_F = G S_{\rm eff}^2 .
\label{ppower}
\end{equation}
When $\theta=0$ or $\pi$, Eqs.~(~\ref{zzt}) and (~\ref{ppower}) give the well-known figure of merit and power factor for the longitudinal thermoelectric effect
\begin{equation}
Z_l T = \frac{G S_1^2 T^2}{K_e^0 - GS_1^2 T^2} , \quad  P_{Fl} = G S_1^2 .
\end{equation}
While the transverse thermoelectric figure of merit and power factor, i.e.,
$\theta=\pi/2$ or $3\pi/2$, are given by
\begin{equation}
Z_t T = \frac{ G S_2^2 T^2}{K_{pe} - G S_2^2 T^2} , \quad P_{Ft} = G S_2^2 .
\end{equation}

\begin{figure}[htb]
\includegraphics[height=10.4cm]{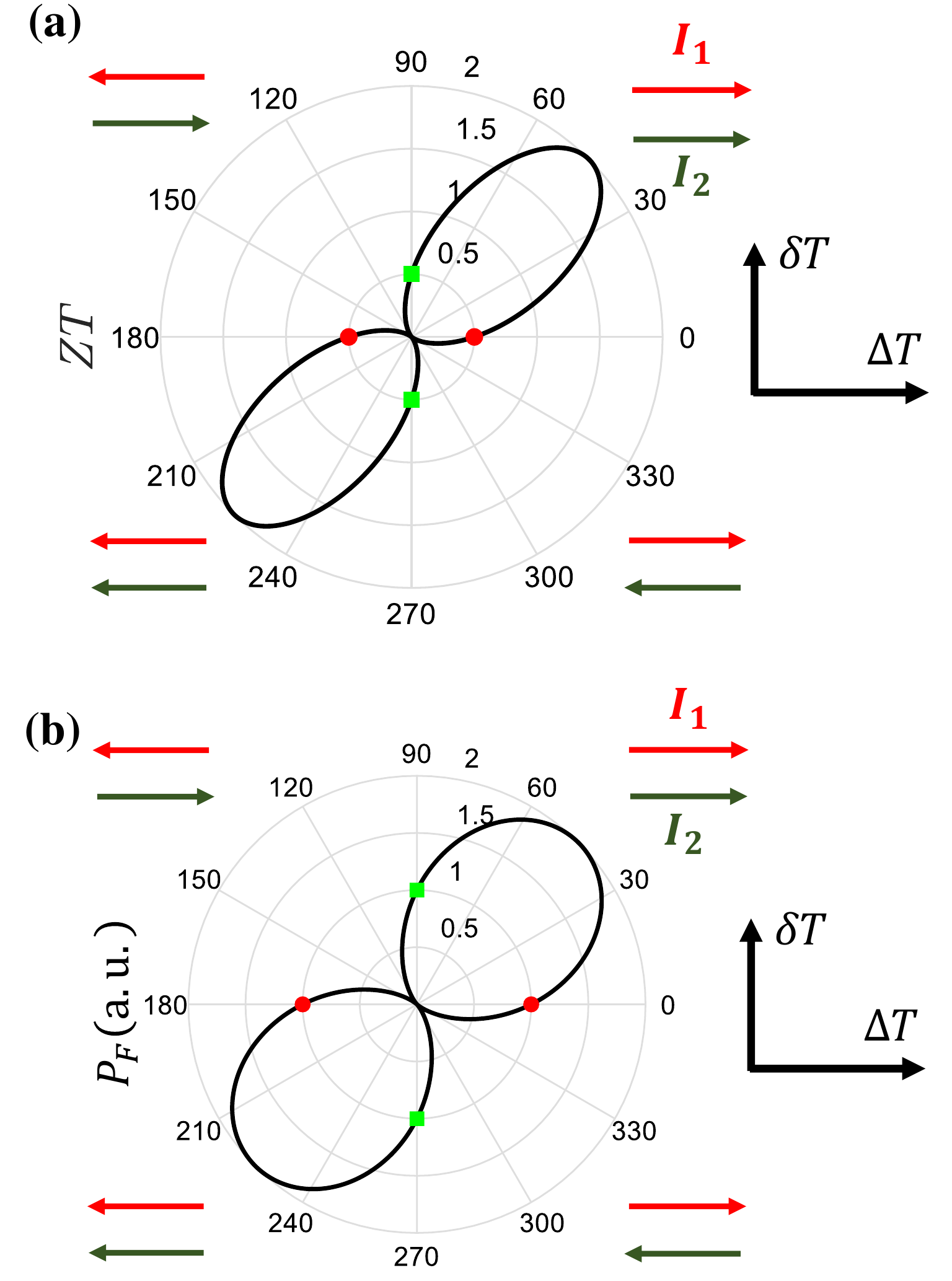}
\caption{Polar plot of (a) figure of merit $ZT$ and (b) power factor $P_F$ [in arbitrary unit (a.u.)] versus angle $\theta$. At $\theta=0^{\circ}$ or $180^{\circ}$ $ZT$ and $P_F$ recover the values for the longitudinal thermoelectric effect (red dots), while at $\theta=90^{\circ}$ and $270^{\circ}$ they go back to those of the transverse thermoelectric effect (green squares). The arrows in the I, II, III, IV quadrants label the direction of the currents $I_1\equiv L_1\Delta T/T$ (red arrows) and $I_2\equiv L_2 \delta T/T$ (green arrows). Figures are reproduced from Ref.~\cite{JiangJAP}.}
\label{JAP_ZT}
\end{figure}

Actually, one can maximize the figure of merit by tuning the angle $\theta$. This is achieved at $\partial_\theta (ZT) = 0$ and one finds the maximum figure of merit is
\begin{equation}
Z_mT = \frac{G(K_e^0K_{pe}-L_3^2)}{ D_{\cal M}} - 1 .
\label{thM}
\end{equation}
where $D_{\cal M}=G K_e^0 K_{pe} - G L_3^2 -  K_{pe} L_1^2 + 2 L_1 L_2 L_3- K_e^0 L_2^2 $ denotes the determinant of the $3\times 3$ transport matrix in Eq.\eqref{eq:Onsager}.
One can also tune $\theta$ to find the maximum power factor
\begin{equation}
P_{Fm} = G (S_1^2+S_2^2)
\end{equation}
is greater than {both} $P_{Fl}$ and $P_{Ft}$ unless $S_1$ or $S_2$ is zero.

Fig.~\ref{JAP_ZT}(a) shows $ZT$ versus the angle $\theta$ in a polar plot for a specific set of transport coefficients. Remarkably for $0<\theta<\pi/2$ and $\pi<\theta<3\pi/2$, $ZT$ is {\em greater} than both $Z_lT$ and $Z_tT$. To understand the underlying physics, we decompose the electric current into three parts $I=I_0+I_1+I_2$ with $I_0\equiv GV$, $I_1\equiv L_1\Delta T/T$, and $I_2=L_2\delta T/T$. The two thermoelectric effects add up constructively as $I_1$ and $I_2$ have the same sign, which takes place when $0<\theta<\pi/2$ and $\pi<\theta<3\pi/2$. Fig. ~\ref{JAP_ZT}(b) shows the power factor versus the angle $\theta$. The power factor is also {\em larger} when the two currents $I_1$ and $I_2$ are in the same direction. Therefore, the cooperation of the two thermoelectric effects leads to an enhanced figure of merit and output power.

Besides the multilayer thermoelectric engines, where one electric current is coupled to two temperature gradients,
the energy cooperation effects in quantum thermoelectric systems with multiple electric currents and only one heat current have also been studied~\cite{MyJAP,CPB}, where the elastic tunneling through quantum dots is considered.
The constructive cooperation in these quantum thermoelectric systems results in the enhanced thermoelectric power and efficiency for various quantum-dot energies, tunneling rates, etc. Moreover, this cooperative enhancement, dubbed as the thermoelectric cooperative effect, is found to be universal in three-terminal thermoelectric energy harvesting~\cite{JordanPRB13,Rongqian}.

\section{Near-field three-terminal thermoelectric heat engine}
Near-field thermal radiation recently emerges as one promising route to efficient transfer heat at the nanoscale~\cite{Zhang2007Nano,song2016,RMPBen}, which dramatically stimulates the advance of thermoelectics~\cite{gchen2005book}.
In Ref.~\cite{JiangNearfield}, we proposed a near-field thermoelectric heat engine, which is composed of two continuous spectra, e.g., narrow-bandgap semiconductor, separately interacting with single quantum dot and inelastically coupled via the near-field thermal emission.
The  near-field inelastic heat engine is exhibited to effectively rectify the charge flow
of photo-carriers and converts {near-field heat radiation} into useful electrical power.
Such near-field thermoelectric device takes the
following advantages of near-field radiations:
First, the near-field radiation can strongly enhance heat transfer across the
vacuum gap and thus leads to significant heat flux injection.
Second, unlike phonon-assisted interband transitions, photon-assisted interband transition is
not limited by the small phonon frequency and can work for
larger band gaps due to the continuous photon spectrum.

\begin{figure}[htb]
\includegraphics[width=8.5cm]{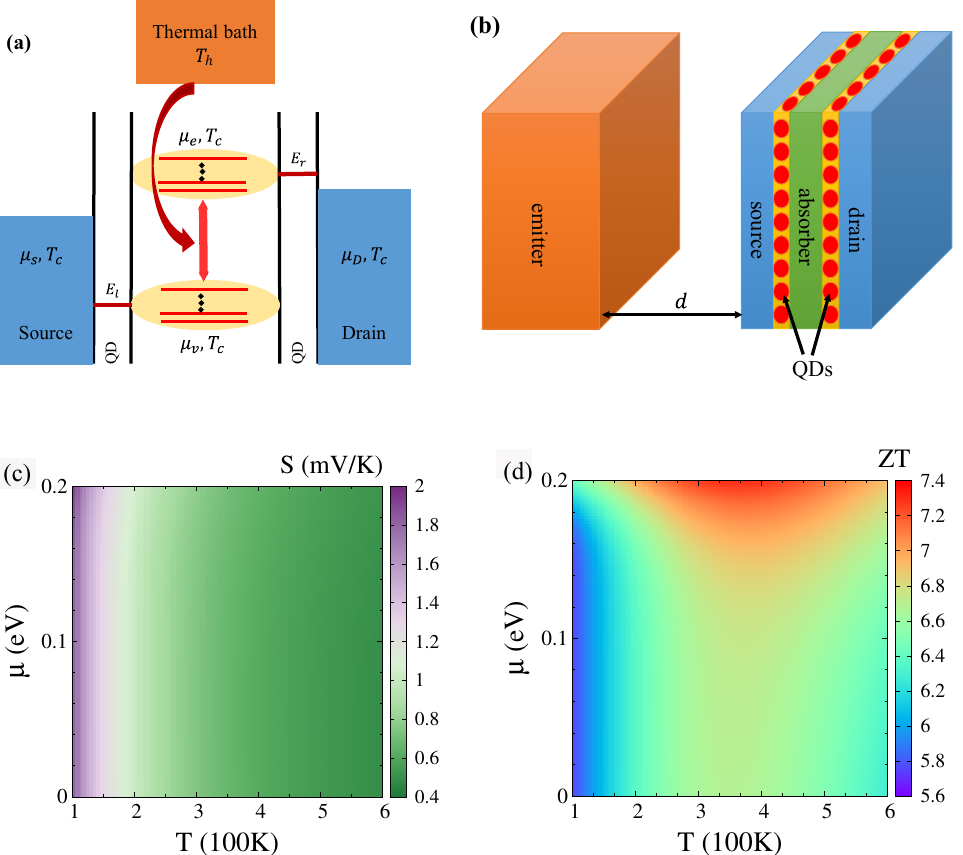}
\caption{(a) Schematic of near-field three-terminal thermoelectric heat engine. A hot thermal reservoir of temperature $T_h$ injects heat flux into the device through near-field heat radiation. The device is held at a lower temperature $T_c$. The absorption of the heat radiation is realized by photon-assisted transitions between the two continua. As a result, the upper and lower continua have different chemical potentials, $\mu_e$ and $\mu_v$, respectively. The source and drain have different electrochemical potentials, denoted as $\mu_S$ and $\mu_D$, separately. The typical energy of QDs in the left (right) layer is $E_{\ell}$ ($E_{r}$).
(b) A possible set-up for the three-terminal near-field heat engine. The emitter is a heat source of temperature $T_h$, which is separated from the device by a vacuum gap of thickness $d$. The device is held at a lower temperature $T_c$ which consists of the source, drain, and absorber layers. These three parts are divided by two layers of quantum dots arrays. (c) Seebeck coefficient $S$ (in unit of mV/K) and (d) thermoelectric figure of merit $ZT$ for the inelastic thermoelectric transport as functions of the chemical potential $\mu$ and the temperature $T$. Figures are reproduced from Ref.~\cite{JiangNearfield}.}
\label{NearField-model}
\end{figure}

Here, we present a microscopic theory for the thermoelectric transport in the near-field inelastic heat engine.
The Hamiltonian of the system is described as
\be
H = H_{\rm SD} + H_{\rm QD} + H_{\rm C} + H_{\rm tun} + H_{\rm e-ph} .
\ee
Specifically, the Hamiltonian for the source and drain is expressed as
$H_{\rm SD} = \sum_{{\vec q}} (E_{S,{\vec q}} c^\dagger_{S,{\vec q}}
c_{S,{\vec q}}  + E_{D,{\vec q}} c^\dagger_{D,{\vec q}}
c_{D,{\vec q}} )$,
where ${\vec q}$ is the wavevector of electrons.
The Hamiltonian of the QDs is
$H_{\rm QD} = \sum_{j=\ell,r} E_{j} d^\dagger_j d_j$,
where $j=\ell,r$ denotes the left and right dot, respectively. We
first consider the case where only one (two if spin degeneracy is
included) level in each QD is relevant for the transport. The
Hamiltonian for the two central continua is
$H_{\rm C} = \sum_{{\vec q}} ( E_{v,{\vec q}} c^\dagger_{v,{\vec q}}
c_{v,{\vec q}}  +  E_{e,{\vec q}} c^\dagger_{e,{\vec q}} c_{e,{\vec
  q}} )$.
The tunnel coupling through the QDs is given by
\begin{align}
H_{\rm tun} =& \sum_{{\vec q}} ( J_{S,{\vec q}} c_{S,{\vec q}}^\dagger
d_\ell + J_{D,{\vec q}} c_{D,{\vec q}}^\dagger
d_r \nn \\
& + J_{v,{\vec q}} c_{v,{\vec q}}^\dagger
d_\ell  + J_{e,{\vec q}} c_{e,{\vec q}}^\dagger
d_r) + {\rm H.c.} .
\end{align}
The coupling coefficients $J$ determine the tunnel rates $\Gamma_{ij}$~\cite{JiangNearfield}.
The Hamiltonian governing the photon-assisted transitions in the center is
\be
H_{\rm e-ph} = \sum_{{\vec q},{\vec k},\tau} \frac{g_{{\vec k},\tau}}{\sqrt{V}} c^\dagger_{e,{\vec q}+{\vec k}} c_{v,{\vec q}} a_{{\vec k},\tau} +
{\rm H.c.},
\ee
where $g_{\vec k}$ is the electron-photon interaction strength, the operator $a_{{\vec k},\tau}$
($\tau=s,p$ denotes the $s$ and $p$ polarized light) annihilates
an infrared photon with polarization $\tau$. $V$ is the volume of the photonic system. 

Via the Fermi golden rule, the thermoelectric transport
coefficients in the linear response regime are obtained as
\begin{subequations}
\begin{align}
& G_{ve} = \frac{e^2}{k_BT}\int d\ome \Gamma_0(\ome) , \\
& L_{ve} = \frac{e}{k_BT}\int d\ome \Gamma_0(\ome) \hbar \ome ,\\
& K_{ve} = \frac{1}{k_BT}\int d\ome \Gamma_0(\ome) \hbar^2 \ome^2,
\end{align}
\label{ave-E}
\end{subequations}
where
\begin{align}
 \Gamma_0(\ome) =& 2\pi \nu_{ph} {\cal F}_{nf} (\ome) \sum_{{\vec q}} |g(\ome)|^2
\delta(E_{e,{\vec q}}-E_{v,{\vec q}}-\hbar\ome) \nn\\
&\times f^0(E_{v,{\vec q}}, T)[1-f^0(E_{e,{\vec q}}, T)]N^0(\ome, T) .  \label{gam0}
\end{align}
The superscript 0 in the above stands for the equilibrium distribution,
$N^0(\ome_{\vec k}, T_c)=1/[\exp(\frac{\hbar\ome_{\vec k}}{k_BT_c})-1]$ is the equilibrium photon distribution function,
and $f^0(E_{v,{\vec q}}, T)=1/[\exp(\frac{E_{v,{\vec q}}-\mu}{k_BT})+1]$ is the Fermi-Dirac distribution function.
$|g(\ome)|^2 = \frac{\hbar\ome d_{cv}^2}{2\vep_0\vep_r}$,
$\nu_{ph}$ is the photon density of states, and the factor is given by
\begin{align}
{\cal F}_{nf}(\ome) = \frac{1}{4} \int_0^{1} \frac{x_kdx_k}{\sqrt{1-x_k^2}} \sum_\tau {\cal T}_{\tau} (\ome, x_k n \ome/c, d),
\end{align}
where $x_k=k_\parallel/(n\ome/c)$.
It is interesting to show that  the photon tunneling probability is specified as~\cite{PRB71,Zhang2007Nano}
\begin{align}
{\cal T}_{\tau}(\ome_{\vec k}, k_\parallel, d) = \left\{ \begin{array}{cccc} \frac{(1-|r^\tau_{01}|^2)(1-|r^\tau_{02}|^2)}{|1-r^\tau_{01}r^\tau_{02}e^{i2k^0_{z}d} |^2}, \quad {\rm if} \quad k_\parallel \le \ome/c \\
\frac{4\Im(r^\tau_{01})\Im(r^\tau_{02})e^{-2\beta^0_z d} }{|1- r^\tau_{01}r^\tau_{02}e^{-2\beta^0_z d} |^2} , \quad {\rm otherwise}
\end{array} \right.
\label{pT}
\end{align}
Here $r^\tau_{01}$ ($r^\tau_{02}$) is the Fresnel reflection coefficient for the interface
between the vacuum (denoted as ``0") and the emitter (absorber) [denoted as ``1" (``2")]. $k^0_z=\sqrt{(\ome/c)^2-k_\parallel^2}$ is
the wavevector perpendicular to the planar interfaces in the vacuum. For $k_\parallel>\ome/c$, the perpendicular wavevector
in the vacuum is imaginary $i\beta^0_z=i\sqrt{k_\parallel^2-(\ome/c)^2}$, where photon tunneling is dominated by evanescent waves.
For isotropic electromagnetic media, the Fresnel coefficients are given by
$r^s_{0j} = \frac{k_z^0-k_z^j}{k_z^0+k_z^j}$ and
$r^p_{0j} = \frac{\vep_jk_z^0 - k_z^j}{\vep_jk_z^0 + k_z^j}~(j=1,2)$,
where $k_z^j=\sqrt{\vep_j(\ome/c)^2-k_\parallel^2}$ and $\vep_j$ ($j=0,1,2$) are the (complex) wavevector along the $z$ direction
and the relative permittivity in the vacuum, emitter, and the absorber, respectively.

Consequently, the Seebeck coefficient of the near-field inelastic three-terminal heat engine is obtained as
\be
S  = \frac{\ave{\hbar\ome}}{eT} ,
\ee
and the figure of merit is given by
\begin{align}
ZT = \frac{\ave{\hbar\ome}^2}{\alpha\ave{\hbar^2\ome^2}-\ave{\hbar\ome}^2+\Lambda_{nf}} , \label{ztnf}
\end{align}
where the average is defined as $\ave{...} = \frac{\int d\ome \Gamma_0(\ome) ... }{\int d\ome \Gamma_0(\ome)}$, $\alpha = {G_{ve}}/{G_{eff}}$, and $\Lambda_{nf}=e^2K_{para}/G_{ve}$ characterizes the parasitic heat conductance $K_{para}$ that does not contribute to thermoelectric energy conversion.
It is shown in Fig.~\ref{NearField-model}(c) that the Seebeck coefficient does not change significantly
by tuning the chemical potential, which is a generic characteristic of the
inelastic thermoelectric effect, for the average energy $\ave{\hbar\ome}$
is mainly limited by the band gap $E_g$ and the temperature $T$.
Moreover, the figure of merit with small parasitic heat conduction,
e.g, $\Lambda_{nf}=0.2E^2_g$ in Fig.~\ref{NearField-model}(d), can be optimized as large as $ZT>7$ around $T=350$~K
and $\mu>0.15$~eV.
Therefore, our work presents one intriguing mechanism of photon-induced inelastic thermoelectricity, which  may
provide physical insight for future thermoelectric technologies based on
inelastic transport mechanisms, and serve as the foundations for future studies.

\section{Quantum efficiency bound for continuous heat engines coupled to non-canonical reservoirs}

The efficiency of heat engines is fundamentally restricted by the second law of thermodynamics to the Carnot limit~\cite{PRXQuantum}. This canonical bound is being challenged nowadays by quantum and classical effects. However, nonequilibrium reservoirs that are characterized by additional parameters besides their temperature are exploited to construct devices with efficiency beyond the Carnot bound~\cite{LutzPRL,KlaersPRX}.

\begin{figure}[htbp]
\centering \includegraphics[width=7.5cm]{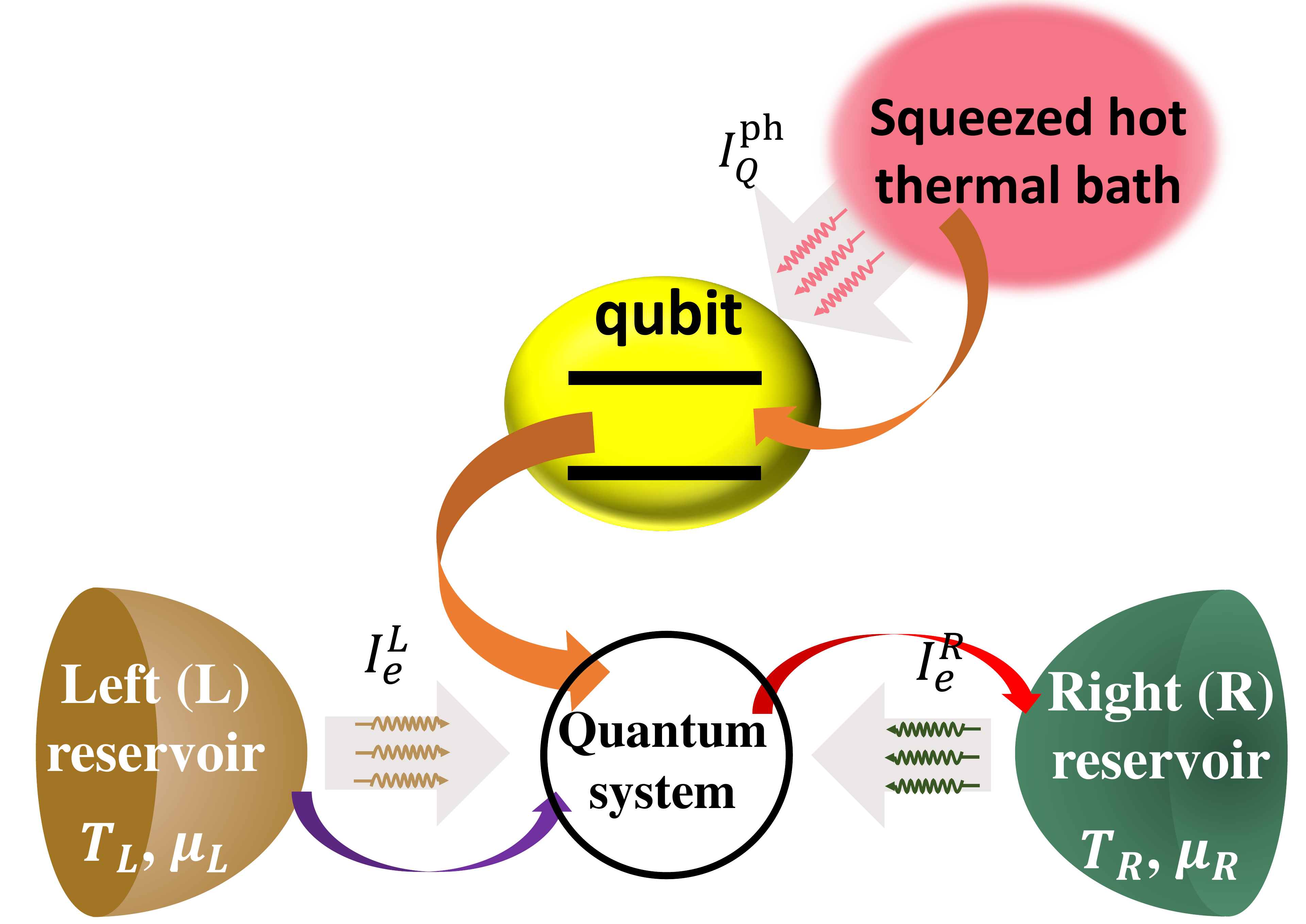}
\caption{Photoelectric quantum heat engine.  Energy absorbed  by the qubit from a hot squeezed thermal reservoir is converted to electrical power in the cold electronic system. }
\label{Bijaysqueezing}
\end{figure}

We study energy conversion in quantum engines absorbing heat from a non-canonical reservoir~\cite{JiangBijayPRB17}. The device consists of a single qubit coupled to hot squeezed photon bath and two cold electronic reservoirs  (the source and drain), shown in Fig. ~\ref{Bijaysqueezing}. In order to describe the system quantum mechanically, we apply the two-time measurement protocol to  define the characteristic function as
\bea
&&\mathcal{Z}(\lambda_c,\lambda_e,\lambda_{\rm ph})
\nonumber\\
&&=\langle e^{i\lambda_c \hat A_c + i\lambda_e \hat A_e + i\lambda_{\rm ph}\hat A_{\rm ph}  }
e^{\!-i\lambda_c \hat A_c(t) \!- i\lambda_e \hat A_e(t) \!- i\lambda_{\rm ph}\hat A_{\rm ph}(t)  }
\rangle.
\nonumber
\label{eq:Z}
\eea
$\lambda_{c,e,{\rm ph}}$ are counting parameters for the charge, electronic energy, and photonic energy, respectively.
$\hat A_c$, $\hat A_e$ and $\hat A_{\rm ph}$ are the respective operators:
$\hat A_c$ is the number operator corresponding to the total charge in the $L/R$ electrode, $\hat A_e$ is the Hamiltonian operator for the $L/R$ electrode and
$\hat A_{\rm ph}$ is the Hamiltonian operator for the photon bath.
$\langle ... \rangle$ represents an average with respect to the total initial density matrix,
which takes a factorized form with respect to the system ($s$) and ($L$, $R$ and $\rm ph$) reservoirs,
$\rho_T(0)=\rho_s(0)\otimes \rho_L\otimes \rho_R\otimes \rho_{\rm ph}$.
The state of the metal leads is described by a grand canonical distribution,
 $\rho_{i}= \exp[-\beta_{\rm el} (\hat H_i-\mu_i \hat N_i)]/Z_i$,
with $Z_i={\rm Tr}\Big[\exp[-\beta_{\rm el} (\hat H_i-\mu_i \hat N_i)]\Big]$
being the partition function, $\beta_i=1/k_BT_i$ being the inverse temperature,
and $\mu_i$ the chemical potential in the $i$th reservoir, respectively.

\subsection{Equilibrium thermal photon bath}

The state of the photon bath is canonical, $\rho_{\rm ph}\!=\! \exp[-\beta_{\rm ph} \hat H_{\rm ph}] / Z_{\rm ph}$, with $Z_{\rm ph}={\rm Tr}\big[\exp(-\beta_{\rm ph} \hat H_{\rm ph})\big]$. The fluctuation relation ${\rm ln} \left[P_t(\Delta S)/P_t(-\Delta S)\right] = \Delta S$ translates to
\bea
\frac{P_t(I_N,I_E,I_Q^{\rm ph})}{P_t(-I_N,-I_E,-I_Q^{\rm ph})} =  e^{\beta_{\rm el} \Delta \mu I_N + (\beta_{\rm el}-\beta_{\rm ph})I_Q^{\rm ph}}.
\eea
Here, $I_N$ denotes the number of electrons transferred from $R$ to $L$ during the time interval $t$. Similarly, $I_E$ is the electronic energy and $I_Q^{\rm ph}$ photonic heat that are exchanged between the baths during the time interval $t$. The characteristics function thus satisfies
\begin{equation}
\begin{aligned}
&\mathcal Z(\lambda_c,\lambda_e,\lambda_{\rm ph}) \\
&= \mathcal Z(-\lambda_c + i\beta_{\rm el}(\mu_R-\mu_L),-\lambda_e,-\lambda_{\rm ph}-i(\beta_{\rm ph}-\beta_{\rm el}))
\label{eq:Zth}
\end{aligned}
\end{equation}
This  relation straightforwardly results in $1= \langle e ^{-\beta_{\rm el} \Delta \mu  I_N + (\beta_{\rm ph}-\beta_{\rm el})I_Q^{\rm ph} }\rangle$.
Using Jensen's inequality, we obtain
$ [-\beta_{\rm el} \Delta \mu \langle I_N \rangle + (\beta_{\rm ph}-\beta_{\rm el})\langle I_Q^{\rm ph}\rangle ] \leq 0$.
Therefore, the efficiency,
$\langle \eta\rangle \equiv -{\Delta \mu \langle I_N\rangle }/{ \langle I_Q^{\rm ph}\rangle}$,
thus obeys the Carnot bound
\bea
\langle \eta \rangle \leq \frac{\beta_{\rm el}-\beta_{\rm ph}}{\beta_{\rm el}}.
\label{eq:eta}
\eea

\subsection{Noncanonical photon bath}

The squeezed thermal reservoir can be depicted as a combination of orthogonal components, which oscillate as $\cos \omega t$ and $\sin \omega t$~\cite{QObook}.
Squeezed states have reduced fluctuations in one of the quadratures---but enhanced noise
in the other quadrature--- to satisfy the bosonic commutation relation.
Such states are defined by two parameters, the squeezing factor $r$ and phase~\cite{QObook}. To restore the detailed balance relation for  the  $r\neq 0$ case,
one can identify an effective temperature~\cite{YiPRE}
\bea
\beta_{\rm eff}= \beta_{\rm ph} + \frac{1}{\hbar\omega_0}\ln\left[ \frac{ 1 + (1+e^{-\beta_{\rm ph}\hbar\omega_0})\sinh^2r   }{ 1 + (1+e^{\beta_{\rm ph}\hbar\omega_0})\sinh^2r} \right],
\label{eq:betaeff}
\eea
which is unique in the present model, with $\hbar\omega_0$ is the energy gap of the qubit.

Identifying the entropy production associated with the photon energy flow by $\langle \Delta S\rangle=(\beta_{\rm el}-\beta_{\rm eff})\langle I_Q^{\rm ph}\rangle$, we  confirm the symmetry Eq. (~\ref{eq:Zth}) by replacing $\beta_{\rm ph}$ with $\beta_{\rm eff}$
\begin{equation}
\begin{aligned}
&\mathcal Z(\lambda_c,\lambda_e,\lambda_{\rm ph})  \\
&= \mathcal Z(\!-\lambda_c \!+ i\beta_{\rm el}(\mu_R\!-\!\mu_L),\!-\lambda_e,\!-\lambda_{\rm ph}\!-i(\beta_{\rm eff}\!-\beta_{\rm el})).
\end{aligned}
\end{equation}
The fluctuation symmetry relation implies that
$1= \langle e ^{-\beta_{\rm el} \Delta \mu I_N + (\beta_{\rm eff}-\beta_{\rm el})I_Q^{\rm ph}}\rangle$.
Thus, the averaged efficiency,
$\langle \eta\rangle \equiv -\Delta \mu\langle I_N\rangle/\langle I_Q^{\rm ph}\rangle $,
is bounded by
\bea
\langle \eta \rangle \leq 1 - \frac{\beta_{\rm eff}}{\beta_{\rm el}}.
\eea
We note that this bound is universal, holding even beyond the squeezed-bath case.
Explicitly, the efficiency bound for our photoelectric engine~\cite{JiangBijayPRB17} is given by
\begin{equation}
\begin{aligned}
\langle \eta \rangle \leq 1 - \frac{\beta_{\rm eff}}{\beta_{\rm el}} +
 \frac{1}{\beta_{\rm el}\hbar\omega_0}\!\ln\left[ \frac{ 1 \!+\! (1+e^{\beta_{\rm ph}\hbar\omega_0})\sinh^2r   }{ 1 \!+\! (1+e^{-\beta_{\rm ph}\hbar\omega_0})\sinh^2r} \right].
\label{eq:boundS}
\end{aligned}
\end{equation}

\begin{figure}
\includegraphics[width=8.5cm]{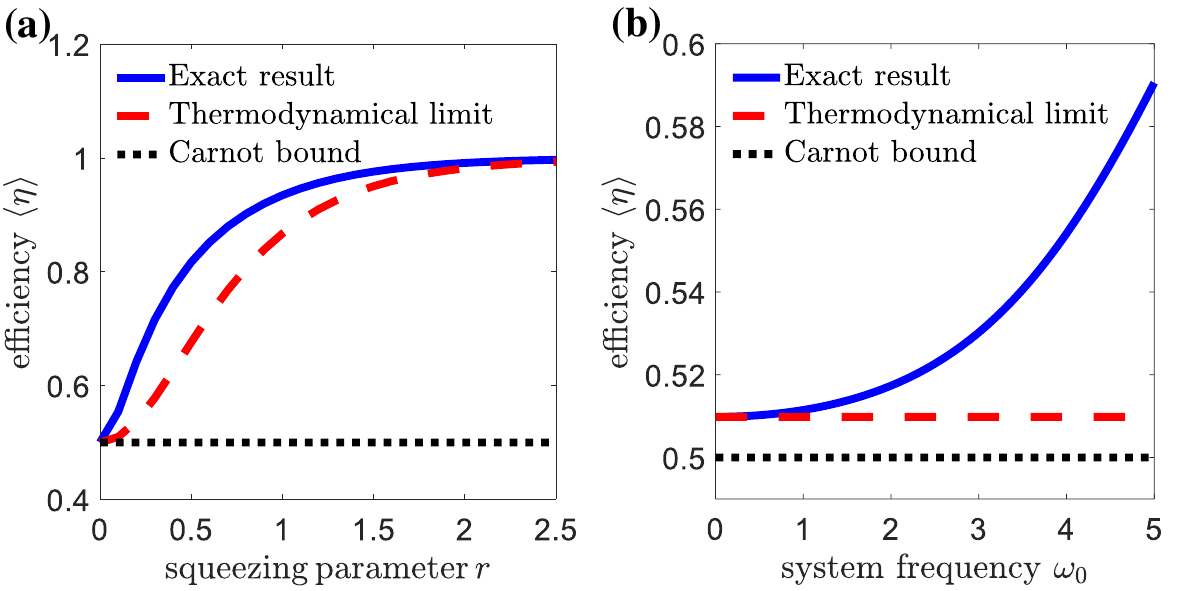}
\caption{Efficiency bound as a function of (a) squeezing parameter $r$ (b) subsystem frequency $\omega_0$. Exact result from Eq. (~\ref{eq:boundS}) (full), thermodynamical limit from Eq. (~\ref{eq:lutz}) (dashed), Carnot bound (dotted). The parameters: $\beta_{\rm el}=2$ and $\beta_{\rm ph}=1$. Figures are reproduced from Ref.~\cite{JiangBijayPRB17}. }
\label{Fig2}
\end{figure}

We now discuss several interesting results of Eq. (~\ref{eq:boundS}). First, we expand it close to thermal equilibrium assuming that $\sinh^2 r$ is a small parameter.
As well, we assume that the temperature of the photon bath is high, e.g., $\beta_{\rm ph}\hbar\omega_0\ll1$.
Then, Eq. (~\ref{eq:boundS}) is reduced to
\bea
\langle \eta \rangle \leq 1- \frac{\beta_{\rm ph}}{\beta_{\rm el} (1+2\sinh^2r)}.
\label{eq:lutz}
\eea
which agrees with Ref.~\cite{LutzPRL14,ParrondoPRE}.
Another interesting case is the deep quantum regime ($\beta_{\rm ph}\hbar\omega_0\gg 1$).
Assuming small $r$, from Eq. (~\ref{eq:boundS}) we receive an exponential quantum enhancement
in comparison to the classical case,
\begin{equation}
\begin{aligned}
\langle \eta\rangle \leq 1-  \frac{\beta_{\rm ph}}{\beta_{\rm el}} + \frac{e^{\beta_{\rm ph}\hbar \omega_0}}{\beta_{\rm el}\hbar \omega_0} \frac{\sinh^2r}{1+\sinh^2r}   .
\end{aligned}
\end{equation}
Fig. ~\ref{Fig2} clearly exhibits these results:
(i) Squeezing enhances the efficiency beyond the Carnot limit.
(ii) In the quantum regime ($\beta_{\rm ph}\omega_0>1$), the bound is greatly reinforced beyond the thermodynamical limit.

\section{Thermoelectric efficiency and its statistics}\label{statistics}

Fluctuations can not be ignored in mesoscopic systems and are particularly important for understanding quantum transport. It can also be considered as a resource for the operations of open quantum systems as functional devices. As a widely used theoretical framework, the fluctuation theorem has been applied to the statistics of the electronic currents, heat currents, and thermodynamic fluctuations~\cite{blickle2012,SeifertPR,CilibertoPRX,Seifert,NP,Verley2014,FT6,QuanPRL18,BijayPRB19,LiuPRE,QuanPRL2,QuanPRL20,MaPRL,FeiPRA,WangJHPRE22}. In this section, from the perspective of statistical physics, we will utilize the fluctuation theorem to analyze thermal fractional devices.

\begin{figure}
\begin{center}
\includegraphics[height=8.0cm]{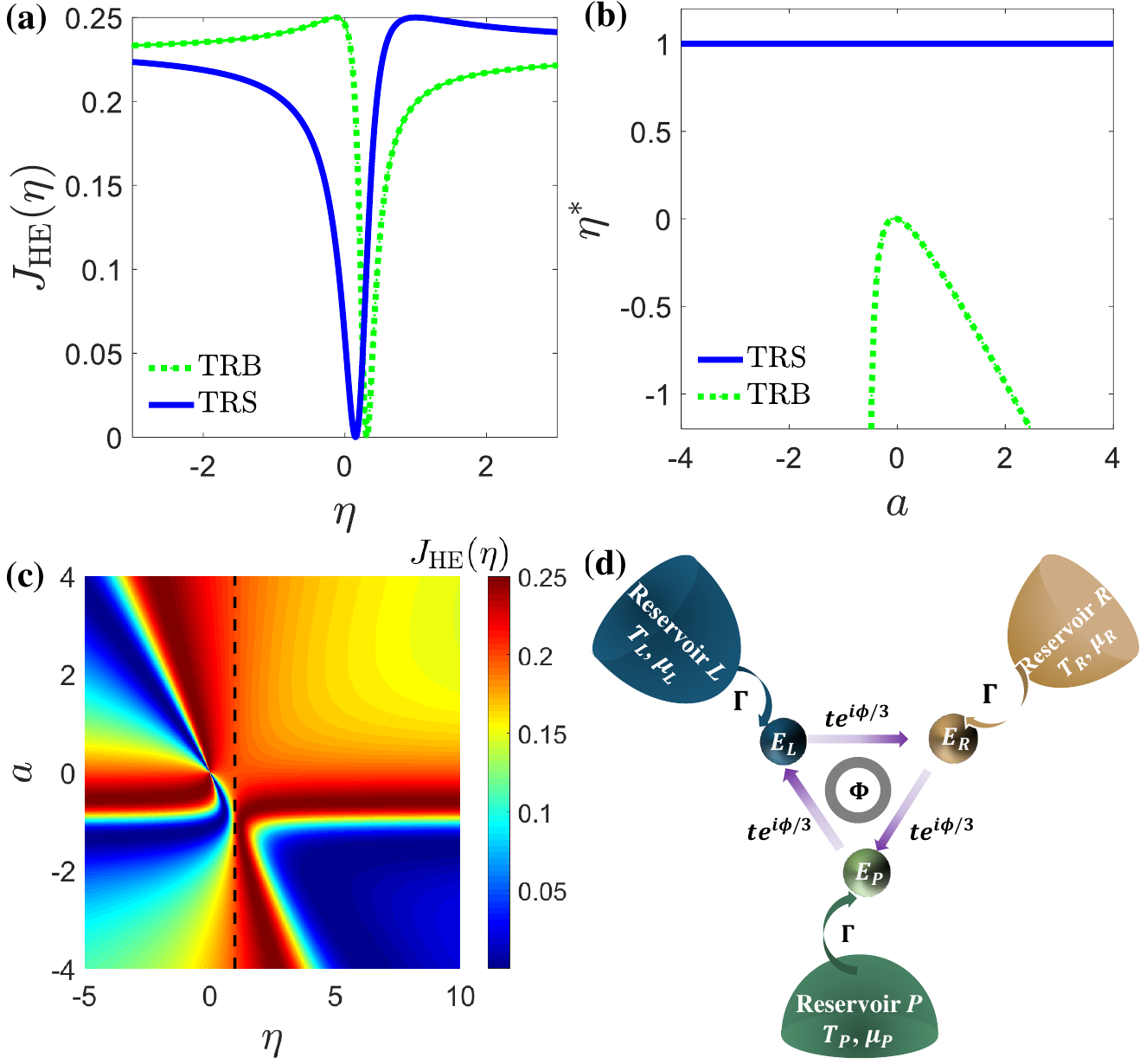}
\caption{Efficiency statistics for TRS and TRB systems. (a) The LDF $J_{\rm HE}(\eta)$ for TRS and TRB cases as function of $\eta$. (b) The least probable efficiency $\eta^\star$ for these two systems at different $a$. (c) LDF $J_{\rm HE}(\eta)$ as a function of $a$ and $\eta$ for a TRB system. (d) A three-terminal triple-QD thermoelectric device with a magnetic flux $\Phi$. Three QD (with $i=1,2,3$) are connected to three electronic reservoirs. The tunneling rates between the QDs and electronic reservoirs is $\Gamma$. Figures are reproduced from Ref.~\cite{JiangPRL}.}
\label{fig:jeta}
\end{center}
\end{figure}

\subsection{Efficiency statistics for three-terminal systems with broken time-reversal symmetry}

By analyzing stochastic efficiency, it was recently shown that the Carnot efficiency is the least likely stochastic efficiency~\cite{FT6}, later found to be solely the consequence of the fluctuation theorem for time-reversal symmetric (TRS) energy transducers~\cite{PDF3}. Breaking the time-reversal symmetry can shift the least likely efficiency away from the Carnot efficiency~\cite{PDF2,PDF3}.

We consider a generic situation in which there are two energy output channels (``1'' and ''2''). Each of the channels has a thermodynamic ``current'' and a affinity. The time-integrated currents are denoted by $J_i$ ($i=1,2$) while the time-intensive current is defined as $I_i=J_i/t$ with $t$ being the total time of operation. A small time-reverse broken (TRB) machine can be characterized in the linear-response regime by $\ov{I}_i=M_{ij}A_j$ ($i,j=1,2$). In this regime the statistics of the currents at long time $t$ can be described within the Gaussian approximation by the distribution $P_t(\vec{I})= \frac{t\sqrt{\det((\hat{M}^{-1})_{sym})}}{4\pi}
\exp(-\frac{t}{4}\delta \vec{I}^T\cdot\hat{M}^{-1}\cdot\delta\vec{I})$~\cite{gaspard,Brownian}. Here $\det((\hat{M}^{-1})_{sym})$ is the determinant of the
symmetric part of the inverse of the Onsager response matrix $\hat{M}$ and the superscript ``$T$'' denotes transpose. The averaged
quantities are represented with a bar over the symbols throughout this
paper. $\delta \vec{I}=\vec{I}-\vec{\ov{I}}$ represents fluctuations
of the currents. From the probability distribution of stochastic
currents we calculate the distribution of efficiency
$P_t(\eta)$. We then obtain the large deviation function
(LDF) of the stochastic efficiency ${\cal G}(\eta)\equiv - \lim_{t\to \infty} t^{-1} \ln[P_t(\eta)] $.

Consequently, the scaled LDF ($J(\eta) \equiv {{\cal G}(\eta)}/{\ov{S}_{tot}}$) is given by
\begin{align}
J_{\rm HE}(\eta)= \frac { J_{\rm HE}(\eta_C) \left ( \eta + a^2 +  \alpha q b + a q
      \eta \right)^2} {(1 + a^2 + a q b + a q)\left (
      \eta^2 + a^2 +  a q\eta +  \alpha q b
      \eta  \right)} ,\label{eq:jeta}
\end{align}
where $\ov{{S}}_{tot}=\sum_i\ov{I}_iA_i$ is the average total entropy production rate and $J_{\rm HE}(\eta_C) = \frac{4-q^2(1+b)^2}{16(1-q^2b)}$ is the scaled LDF at Carnot efficiency. Here, $q = \frac{M_{21}}{\sqrt{M_{22}M_{11}}}$, $b = \frac{M_{12}}{M_{21}}$, $a = \frac{A_1\sqrt{M_{11}}}{A_2\sqrt{M_{22}}}$ are dimensionless parameters that characterize the responses of the system and the applied affinities. In our scheme, efficiency is scaled so that the Carnot (reversible) efficiency corresponds to $\eta_C\equiv 1$.

In particular, the minimum $J_{\rm HE}(\bar\eta_{\rm HE})=0$ is reached at the average efficiency ${\bar\eta}_{\rm HE}=-a(a+qb)/(a q+1)$,
whereas the maximum value $J_{\rm HE}(\eta^\star)=1/4$ is realized at the least probable efficiency
\be
\eta^\star = 1 + \frac{q (b -1) (1 + a q + a q b + a^2 )}{q - q b - 2 a + q^2 (1 + b) a}.
\label{eta-star}
\ee
In the TRS limit, the least likely efficiency is  always identical to the Carnot efficiency, $\eta^{\rm \star}=\eta_C\equiv 1$. For TRB systems, in contrast, we find here that $\eta^\star$ {\em depends on} the parameters $q$, $a$, and $b$, see Figs.~\ref{fig:jeta}(a), ~\ref{fig:jeta}(b), and ~\ref{fig:jeta}(c).

\begin{figure}[htbp]
\centering\includegraphics[width=8.5cm]{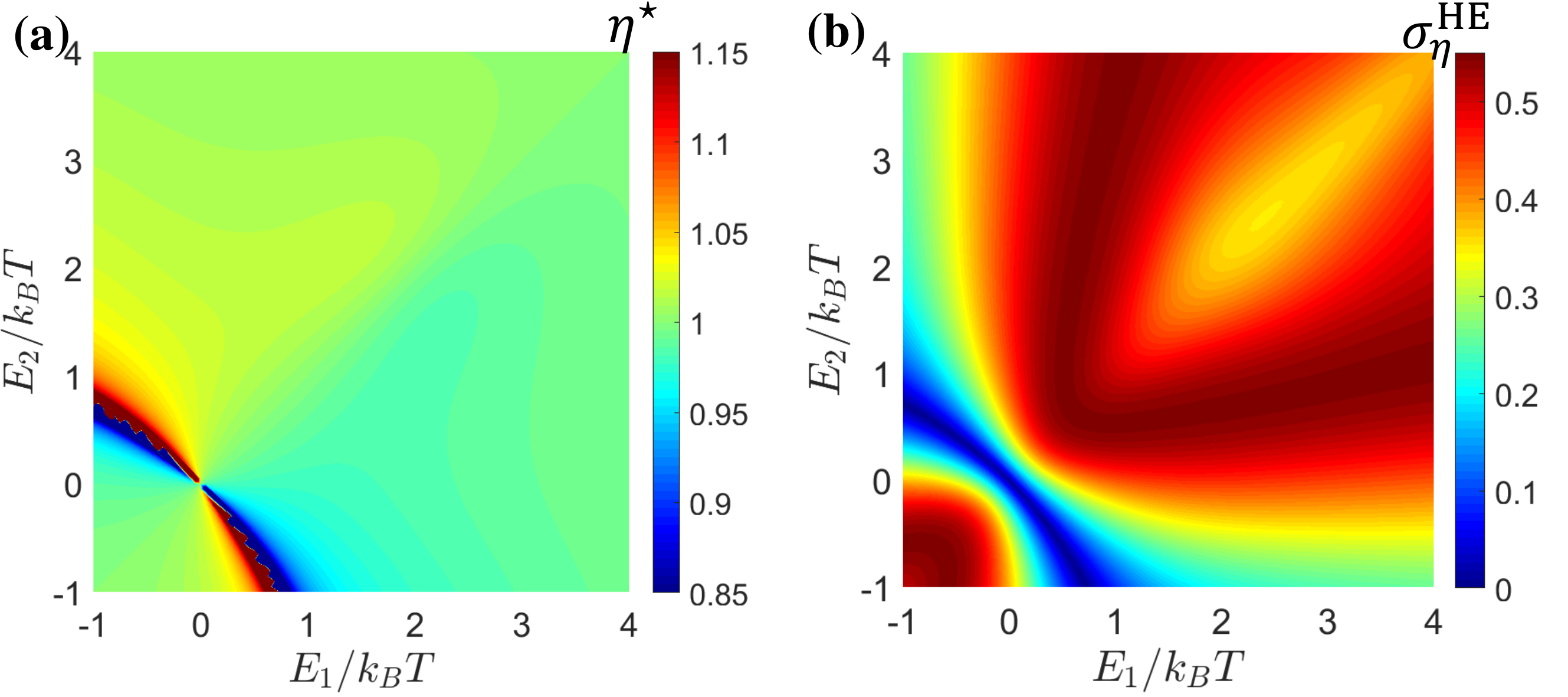}
\caption{Triple-QD thermoelectric system at maximum average output power condition: (a) Least probable efficiency $\eta^{\star}$, (b) The width of efficiency distribution $\sigma_\eta^{\rm HE}$ as functions of QD energies. The white region in (a) depicts very large or very small (negative) $\eta^{\star}$ values which are not properly displayed. Figures are reproduced from Ref.~\cite{JiangPRL}.} \label{fig:3QD-eta}
\end{figure}

Moreover, the width of the distribution around the average efficiency, $\sigma_\eta^{\rm HE}$, is considered as another key characteristic of efficiency fluctuations. Expanding $J_{\rm HE}(\eta)$ around its minimum $\ov{\eta}_{\rm HE}$, one writes $J_{\rm HE}(\eta)\simeq \frac{1}{2(\sigma^{\rm HE}_{\eta})^2}(\eta - \ov{\eta}_{\rm HE})^2 + {\cal O}((\eta-\ov{\eta}_{\rm HE})^3)$, to provide here
\be
\sigma_\eta^{\rm HE} = \frac{2 \sqrt{2} |a| (1 - q^2 b) (1 + a^2 + a q + a b q)}{(1 + a q)^2 \sqrt{4 - q^2 (1 + b)^2}}.
\label{sig}
\ee
We exemplify our analysis within a mesoscopic triple-QD thermoelectric device under a piercing magnetic flux, as shown in Fig. ~\ref{fig:3QD-eta}.

\subsection{Large-deviation function for efficiency: beyond linear-response}

In the following, we study the statistics of efficiency fluctuations in the non-equilibrium regime.
In a recent study, Esposito {\it et al.} analyzed the thermoelectric efficiency statistics in a purely coherent charge transport model~\cite{FT3}.
In parallel, classical models were also examined~\cite{Verley2014}.
Alternatively, the three-terminal device offers a rich opportunity to examine the thermoelectric efficiency beyond linear response, explore the new concept of efficiency fluctuations, and interrogate the role of quantum effects and many-body interactions on the operation of a molecular thermoelectric engine.

\begin{figure}[htb]
\begin{center}
\centering \includegraphics[width=6.5cm]{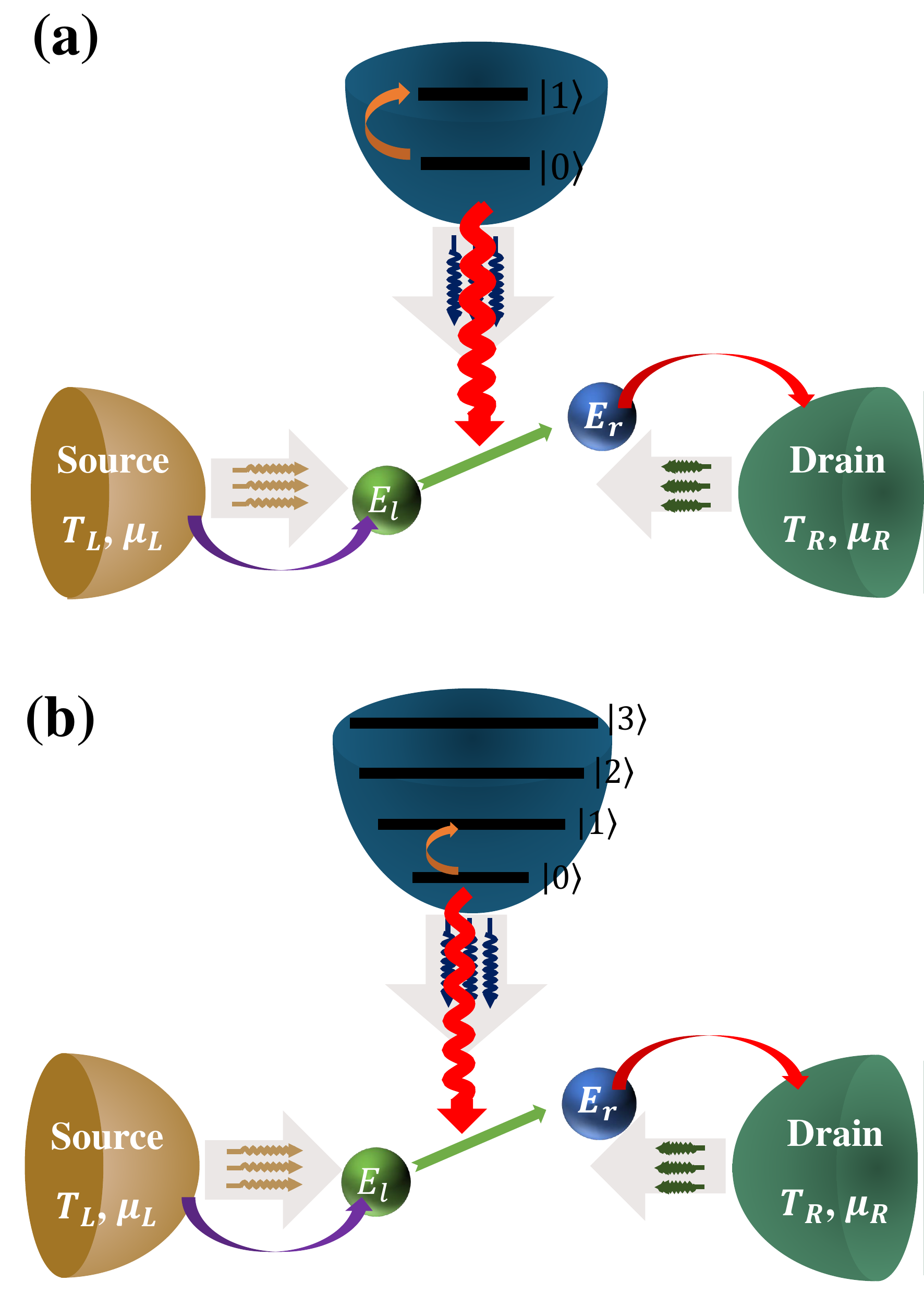}
\caption{Scheme of a three-terminal QD system. (a) The electron transfer is coupled to a highly anharmonic impurity mode which consists of two QD with $E_l$ and $E_r$. (b) The vibrational mode is assumed harmonic. }
\label{BijayPRB}
\end{center}
\end{figure}

\begin{figure}[htb]
\begin{center}
\centering \includegraphics[width=8.5cm]{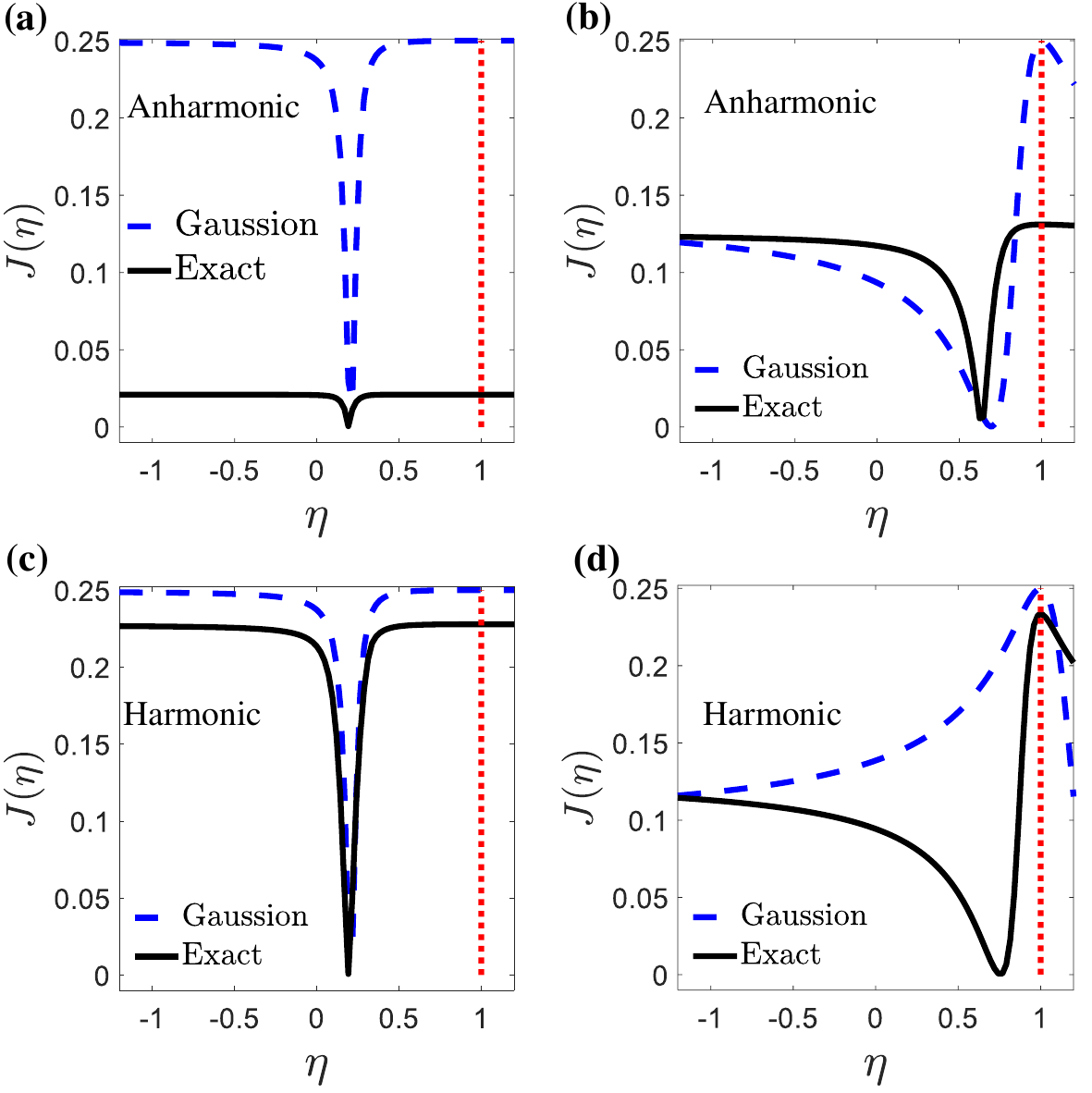}
\caption{(a) and (b) efficiency LDF $J_{\rm NL}(\eta)$ for the anharmonicity vibrational mode model with different bias voltage (a) $\Delta\mu=0.025\, \rm eV$, (b) $\Delta\mu=0.1 \,\rm eV$ for Gaussian limit [Eq. \eqref{eq:jeta}] and exact solution [Eq. \eqref{eq:tildeJ}]. (c) and (d) efficiency LDF $J_{\rm NL}(\eta)$ for the harmonicity vibrational mode model with different bias voltage (a) $\Delta\mu=0.025\, \rm eV$, (b) $\Delta\mu=0.1 \,\rm eV$ for Gaussian limit [Eq. \eqref{eq:jeta}] and exact solution [Eq. \eqref{eq:tildeJ}]. Figures are reproduced from Ref.~\cite{BijayJiang}. }
\label{BijayPRB}
\end{center}
\end{figure}

Due to the stochastic nature of small systems, efficiency fluctuations are typically not bounded and can take arbitrary values. In general, it is useful to investigate the probability distribution function $P_t(\eta)$ to obtain the fluctuating work and heat within the interval $t$, also to observe the value $\eta$ within time $t$. According to the theory of large deviations, the probability function assumes an asymptotic long time form~\cite{Touchette,AgarwallaBJN},
\be
P_t(\eta) \sim e^{-t J_{\rm NL}(\eta)}
\ee
with $J_{\rm NL} (\eta)$ being the ``large deviation function". The large deviation function for efficiency can be obtained from
${\cal G}(\lambda_w, \lambda_q)$ by setting $\lambda_q=\eta \,\eta_C \lambda_w$, and minimizing it with respect to
$\lambda_w$,
\be
J_{\rm NL}(\eta) = - \min_{\lambda_w} {\cal G}(\lambda_w, \eta\, \eta_C\, \lambda_w).
\label{eq:tildeJ}
\ee
where ${\cal G}(\lambda)$ is the cumulant generating function (CGF) of the three-terminal device. $\lambda_w$ and $\lambda_q$ are the counting fields for work and heat, respectively. Note that we do not explicitly evaluate the probability distribution function $P_t(\eta)$. It can be confirmed that $J_{\rm NL}(\eta)$ has a single minimum, coinciding with the macroscopic efficiency of the engine, and a single maximum, corresponding to the least likely efficiency, which equals to the Carnot efficiency, i.e., $\eta=1$.

We numerically investigate the thermoelectric efficiency and its statistics in the three-terminal device, considering the effects of mode anharmonicity and harmonic vibrational mode beyond linear-response situations. The CGFs for an anharmonic impurity and harmonic vibrational mode models are given by Ref.~\cite{BijayJiang}. In Fig.~\ref{BijayPRB} we compare the scaled LDF $J(\eta)$ for the two modes in linear-response regime [Eq. \eqref{eq:jeta}] and beyond linear-response regime [Eq. \eqref{eq:tildeJ}].
It is found that the position $\eta$ for the minimum of $J(\eta)$ can be well captured based on the Gaussian assumption in the linear-response regime.
Moreover, such coincidence also persists even at finite thermodynamic bias for the anharmonic case.

\subsection{Brownian linear thermal transistors }
In Ref.~\cite{MyPRBtransistor}, we have studied the statistical distributions of thermal transistor amplification factor and the cooling by heating efficiency under the assumption of the Gaussian fluctuation.
Particularly in the linear-response regime, the statistics of the stochastic heat currents at long time can be described within the Gaussian approximation~\cite{PDF2} by the distribution $P_i(Q_L^{(i)},Q_{\rm ph}^{(i)}) = \frac{t\sqrt{\det(\hat{K}^{-1})}}{4\pi}\exp[-\frac{t}{4}\Delta\vec{Q}_i^T\cdot\hat{K}^{-1}\cdot\Delta\vec{Q}_i]$. Here  $Q_{L(\rm ph)}^{(1)}=I_Q^{L(\rm ph)}(T+\delta T)$ and $Q_{L(\rm ph)}^{(2)}=I_Q^{L(\rm ph)}(T_{\rm ph}=T)$ with $\delta T/T\rightarrow0$. $\Delta\vec{Q}=\vec{Q}-\vec{\overline{Q}}$ represents fluctuations of the heat currents, where $\vec{\overline{Q}}$ is the average heat current and the $\vec{Q}$ is the stochastic one. From the probability distribution of stochastic heat currents, we have the LDF of stochastic thermal transistor~\cite{MyPRBtransistor}
\begin{equation}
h(\alpha)=\frac{[(K_{12}- K_{22}\alpha)\Delta A_{\rm ph}]^2}{8(K_{11}-2 K_{12}\alpha+K_{22}\alpha^2)}.
~\label{eq:hxi}
\end{equation}
where $\Delta A_{\rm ph}=A_{\rm ph}^{(1)}-A_{\rm ph}^{(2)}$ and $A_{\rm ph(L)}^{(i)}$ ($i=1,2$) are the affinities for heat currents $Q_{\rm ph(L)}^{(i)}$.

The minimum $h(\bar\alpha)=0$ locates at the average transistor amplification~\cite{Jiangtransistors}
\begin{equation}~\label{ampeff1}
\overline\alpha=\frac{K_{12}}{K_{22}},
\end{equation}
which correspond to the maximal probability for the appearance of the amplification efficiency. 

The amplification fluctuation is obtained as
\begin{equation}
\sigma_{\alpha}=\frac{2\sqrt{K_{22}(K_{11}K_{22}-K_{12}^2)}}{K_{22}^2\Delta A_{\rm ph}},
\end{equation}
which obeys the bound of the Onsager coefficients $K_{11}K_{22}-K_{12}^2\ge0$ and $K_{22}{\ge}0$~\cite{JiangPRE}. The equality is reached as the fluctuation width completely vanishes. Obviously, when this equality is reached, the total entropy production rate of the system in the linear-response regime is $dS/dt\equiv0$, i.e., the system is in the equilibrium state.

\begin{figure}[htb]
\begin{center}
\centering \includegraphics[width=8.5cm]{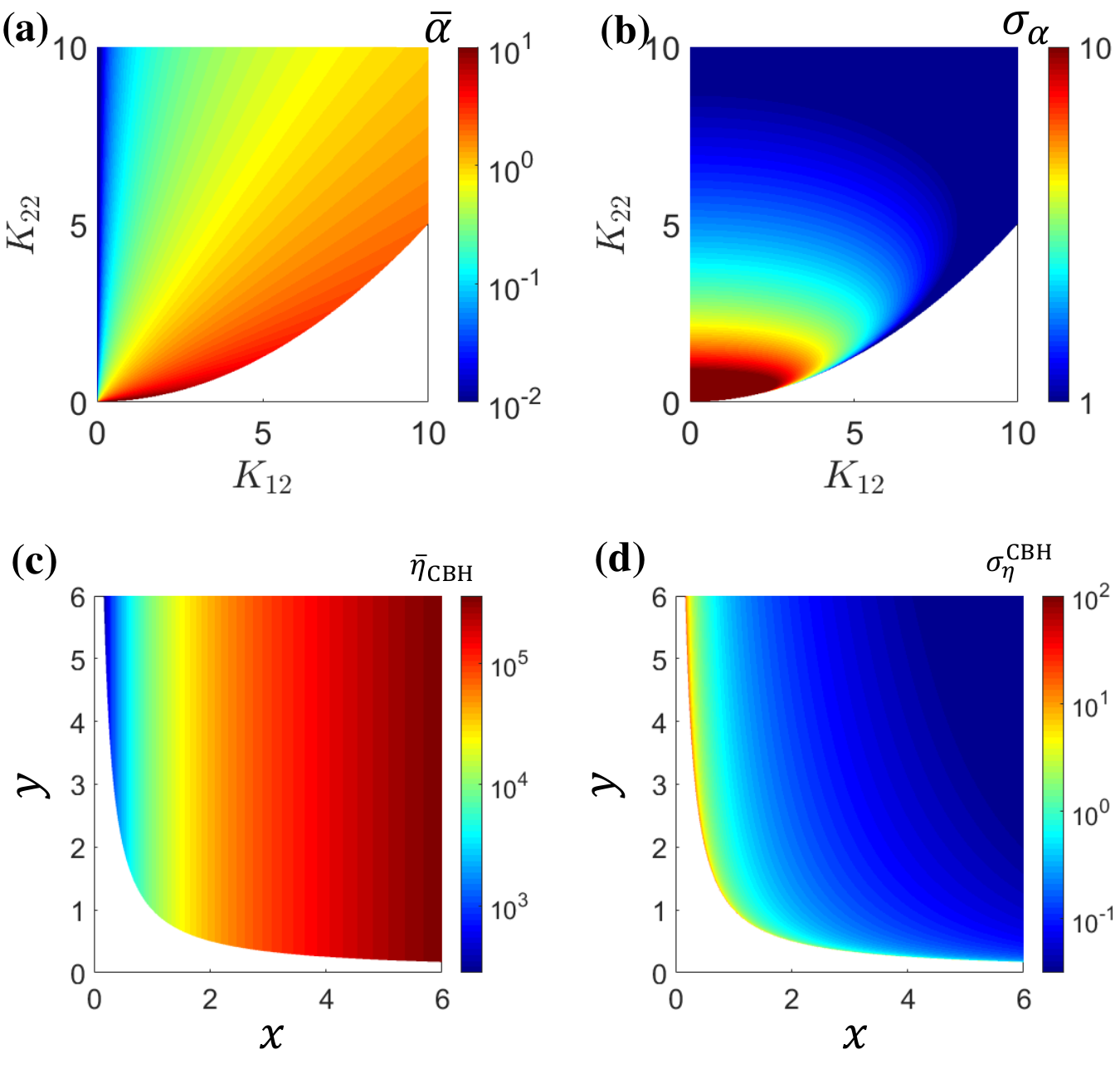}
\caption{(a) $\bar\alpha$ and (b) $\sigma_{\alpha}$ as functions of $K_{12}$ and $K_{22}$. The white region is forbidden by the thermodynamic bound. (c) The average efficiency $\bar\eta_{\rm CBH}$ and (d) the width of cooling efficiency distribution $\sigma_\eta^{\rm CBH}$. The white region is forbidden by the thermodynamic bound. Figures are reproduced from Ref.~\cite{MyPRBtransistor}. }\label{fig:cooling}
\end{center}
\end{figure}

\subsection{The statistics of refrigeration efficiency}


Here, we reveal the fluctuations of cooling by heating refrigerators in the linear-response regime. The scaled LDF of stochastic efficiency~\cite{FT6,FT3} can be expressed as
\begin{equation}
g_{\rm CBH}(\eta)=\frac{[1-y\eta+(x-\eta)z]^2}{4[x+\eta(-2+y\eta)](x z^2+2z+y)},
\label{eq:geta}
\end{equation}
with dimensionless parameters $x={K_{11}}/{K_{12}}$, $y={K_{22}}/{K_{12}}$, and $z={A_L}/{A_{\rm ph}}$. The thermodynamic forces are $A_L=(T_R-T_L)/T$ and $A_{\rm ph}=(T_{\rm ph}-T_L)/T$, respectively.

The minimum of $g_{\rm CBH}(\bar\eta_{\rm CBH})=0$ is reached at the average efficiency
\begin{equation}
\bar\eta_{\rm CBH}=\frac{xz+1}{y+z}.
\end{equation}

The fluctuating width of the average efficiency, $\sigma_{\eta}^{\rm CBH}$, is obtained by expanding $h(\overline\eta_{\rm CBH})=0$ around its minimum $\overline\eta_{\rm CBH}$,
\begin{equation}
\sigma_\eta^{\rm CBH}=\frac{(y+z)^2}{(x z^2+2z+y)\sqrt{2(xy-1)}}.
\end{equation}
Figs.~\ref{fig:cooling}(c) and ~\ref{fig:cooling}(d) illustrate the cooling efficiency $\bar\eta_{\rm CBH}$ and the behavior of the width of cooling efficiency distribution $\sigma_\eta^{\rm CBH}$ when $z=\infty$. We can observe that the $\sigma_{\eta}^{\rm CBH}$ reaches the maximum under the limit condition, i.e., $(xz+1)(y+z)=0$.

In summary of this section, we emphasize that the statistics of energy efficiency can reveal information on the three-terminal thermoelectric system in the linear and nonlinear regime, and the average efficiency and its fluctuations can further characterize the properties of the system.

\section{Thermophotovolatic systems based on near-field tunneling effect}\label{NTPV}

As a solid-state renewable energe resource, thermophotovoltaic (TPV) systems have immense potentials in a wide range of applications including solar energy harvesting and waste heat recovery~\cite{liao2016efficiently,zhao2017high,Tervo2018}.
In the TPV system, a photovoltaic (PV) cell is placed in the proximity of a thermal emitter and converts the thermal radiation from the emitter into electricity via infrared photoelectric conversion. However,  the TPV performance is significantly reduced due to the frequency mismatch between the thermal emitter and the PV cell in the TPV systems at moderate temperatures (i.e., 400$\sim$900~K which is the majority spectrum of the industry waste heat).
To overcome this obstacle, materials which support surface polaritons have been used to introduce a resonant near-field energy exchange between the emitter and the absorber~\cite{ilic2012overcoming,svetovoy2014graphene,zhao2017high}. As a consequence, near-field TPV (NTPV) systems have been proposed to achieve appealing energy efficiency and output power~\cite{laroche2006JAP,molesky2015ideal}. Near-field systems based on graphene, hexagonal-boron-nitride ({\it h}-BN) and their heterostructures have been shown to demonstrate excellent near-field couplings due to surface plasmon polaritons (SPPs), surface phonon polaritons (SPhPs) and their hybridizations [i.e., surface plasmon-phonon polaritons (SPPPs)]~\cite{svetovoy2012plasmon,messina2013graphene,Bo_JHT,Bo_PRB,Sailing_ACS}. In Ref.~\cite{PRAppliedWang}, we propose to use graphene-{\it h}-BN heterostructures~\cite{SPPPs1,SPPPs2,Bo_JHT,Bo_PRB,Sailing_ACS} as the emitter and the graphene-covered InSb $p$-$n$ junction as the TPV cell. We find that such a design leads to significantly
improved performance as compared to the existing studies~\cite{messina2013graphene,heavens1991optical,knittl1976optics}.

\begin{figure}
\begin{center}
\includegraphics[width=3.2 in,height=1.8 in]{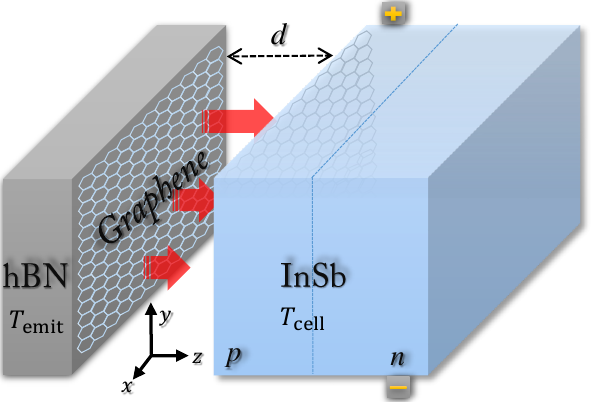}
\caption{\ Schematic representation of the near-field thermophotovoltaic system. A thermal emitter of temperature $T_{\rm emit}$ made of {\it h}-BN/graphene heterostructure is placed in the proximity of a thermophotovoltaic cell of temperature $T_{\rm cell}$ made of InSb. The emitter-cell distance is kept at $d$. The red arrows represents the heat flux radiated from the emitter to the cell. The coordinate axes on the left side shows the in-plane (parallel to the $x$-$y$ plane) and out-of plane (perpendicular to the $x$-$y$ plane) directions. Figures are reproduced from Ref.~\cite{PRAppliedWang}.}
\label{nfTPV-model}
\end{center}
\end{figure}

Fig. ~\ref{nfTPV-model} presents the proposed NTPV system. The emitter is a graphene covered {\it h}-BN film of thickness $h$, kept at temperature
$T_{\rm emit}$. The TPV cell is made of an InSb $p$-$n$ junction, kept at temperature $T_{\rm cell}$, which is also covered by a layer of graphene. The thermal radiation from the emitter is absorbed by the cell and then converted into electricity via photoelectric conversion. The performance of the NTPV system is characterized by the output electric power density  $P_{\rm e}$ and energy efficiency $\eta$.

The output electric power density $P_{\rm e}$ of the NTPV system is defined as the product of the net electric current density  $I^{\rm NTPV}_{\rm e}$ and the voltage bias $V$~\cite{PRAppliedWang},
\begin{equation}
	\begin{aligned}
		P_{\rm e}^{\rm NTPV}= -I^{\rm NTPV}_{\rm e}V,
		\label{epower}
	\end{aligned}
\end{equation}
and the energy efficiency $\eta^{\rm NTPV}$ is given by the ratio between the output electric power density $P^{\rm NTPV}_{\rm e}$ and incident radiative heat flux $Q_{\rm inc}$,
\begin{equation}
	\begin{aligned}
		\eta^{\rm NTPV}= \frac{P_{\rm e}^{\rm NTPV}}{Q_{\rm inc}},\label{effi}
	\end{aligned}
\end{equation}

The incident radiative heat flux is given by
\begin{align}
	Q_{\rm rad}= Q_{\omega<\omega_{\rm gap}} + Q_{\omega\ge\omega_{\rm gap}} \label{Qrad}
\end{align}
where $Q_{\omega<\omega_{\rm gap}}$ and $Q_{\omega\ge\omega_{\rm gap}}$ are the heat exchanges below and above the band gap of the cell, respectively~\cite{polder1971theory,pendry1999radiative}.

The electric current density of a NTPV cell is calculated via the detailed balance analysis~\cite{shockley1961detailed},
\begin{equation}
	\begin{aligned}
		I=I_{\rm ph}-I_0[\exp(V/V_{\rm cell})-1],  \label{Ie}
	\end{aligned}
\end{equation}
Where $V_{\rm cell}=k_{\rm B}T_{\rm cell}/e$ is a voltage which measures the temperature of the cell~\cite{shockley1961detailed}. $I_{\rm ph}$ and $I_0$ are the photo-generation current density and reverse saturation current density, respectively. The reverse saturation current density is determined by the diffusion of minority carriers in the InSb $p$-$n$ junction and the photo-generation current density $I_{\rm ph}$ is contributed from the above-gap thermal heat exchange~\cite{PRAppliedWang}.

\begin{figure}
\centering\includegraphics[width=8.5cm]{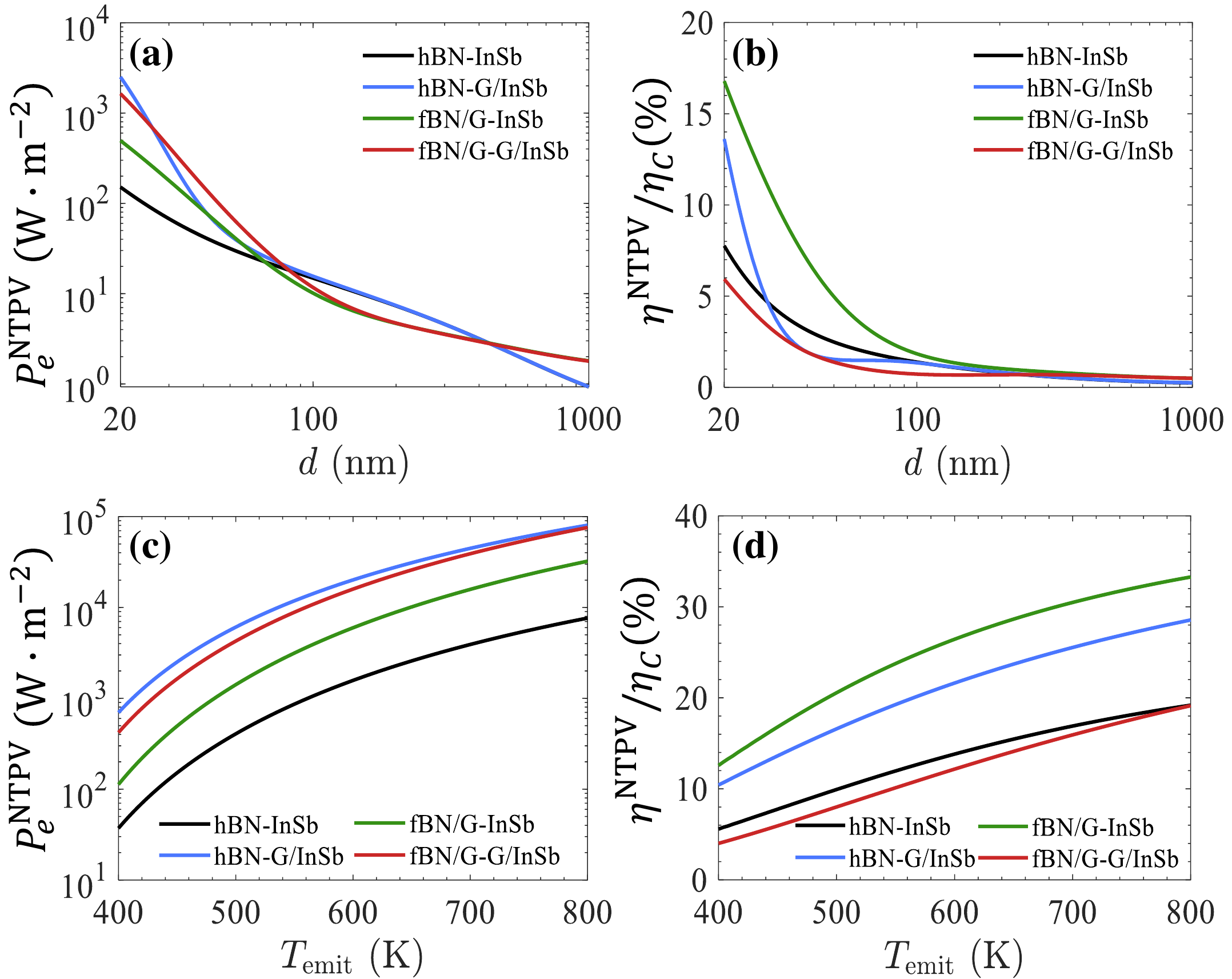}
\caption{Optimal performances of the four NTPV devices. (a) and (b), Optimal (a) output power density $P^{\rm NTPV}_{\rm e}$ and (b) energy efficiency in unit of the Carnot efficiency ($\eta ^{\rm NTPV}/ \eta_{\rm C}$) as functions of the vacuum gap $d$. The temperatures of the emitter and the cell are set as $T_{\rm emit}=450$~K and $T_{\rm cell}=320$~K, respectively. (c) and (d), Optimal (c) output power density $P^{\rm NTPV}_{\rm e}$ and (d) energy efficiency in unit of the Carnot efficiency ($\eta^{\rm NTPV} / \eta_{\rm C}$) as functions of the emitter temperature $T_{\rm emit}$ with $d=20$~nm and $T_{\rm cell}=320$~K. For all these figures, the chemical potential of graphene is set as $\mu_{\rm g}=1.0$~eV. The chemical potential difference across the InSb $p$-$n$ junction $\Delta \mu$ is optimized independently for each configuration. Figures are reproduced from Ref.~\cite{PRAppliedWang}}\label{fig:performance}
\end{figure}

The performances of four different NTPV configurations are examined: (i) the $h$-BN-InSb device (denoted as hBN-InSb,
with the mono-structure bulk $h$-BN being the emitter and the uncovered InSb $p$-$n$ junction being the cell), (ii) the $h$-BN-graphene/InSb
device (denoted as hBN-G/InSb, with the bulk $h$-BN being the emitter and the graphene-covered InSb $p$-$n$ junction as the cell), (iii) the
$h$-BN/graphene-InSb device (denoted as fBN/G-InSb, with the $h$-BN/graphene heterostructure film being the emitter and the
uncovered InSb $p$-$n$ junction as the cell), and (iv) the $h$-BN/graphene-graphene/InSb device (denoted as fBN/G-G/InSb, with the $h$-BN/graphene
heterostructure film being the emitter and the graphene-covered InSb $p$-$n$ junction as the cell). We study and compare their
performances for various conditions to optimize the performance of the NTPV system. As shown in Fig.~\ref{fig:performance},
the primitive hBN-InSb set-up has poor energy efficiency and output power.

The overall best performance comes from the $h$BN-G/InSb (if high output power is preferred) and the fBN/G-InSb (if high energy efficiency
is preferred) set-ups. The underlying physics for the different characteristics of the four different set-ups is understood as due to the resonant coupling between the emitter and the $p$-$n$ junction, where the SPPs in graphene and SPhPs in $h$-BN play crucial roles~\cite{PRAppliedWang,Bo_JHT,Bo_PRB,Sailing_ACS}.

Since the semiconductor thin-films have been explored in NTPV systems, we further investigate the performance of the NTPV systems based on  thin-film $p$-$n$ junctions. A NTPV system based on a InAs thin-film cell with appealing performance operating at high temperatures has been recently proposed ~\cite{zhao2017high,papadakis2020broadening}. But the system suffers from low energy efficiency (below $10\%$) when operating at moderate temperature due to the parasitic heat transfer induced by the phonon-polaritons of InAs. In Ref.~\cite{CPL}, we use InSb as the near-field absorber since the bandgap energy of InSb is lower compared to InAs and its photon-phonon interaction is much weaker than InAs. In this work, we examine the performances of two NTPV devices: the graphene-$h$-BN-graphene-InSb cell (denoted as G-FBN-G-InSb cell, with the graphene-$h$-BN-graphene sandwich structure being the emitter and the InSb thin-film being the cell) and the graphene-$h$-BN-graphene-$h$-BN-InSb cell (denoted as G-FBN-G-FBN-InSb cell, with the double graphene-$h$-BN heterostructure being the emitter and the InSb thin-film being the cell). It is found that the G-FBN-G-InSb cell, despite having a simpler structure, performs better than the G-FBN-G-FBN-InSb cell. While both of the NTPV systems based on InSb thin-film cell underperform the ones based on bulk InSb cell. This is due to the exponential decay characteristic of the electromagnetic wave propagating in the InSb thin-film, which induces an actual availability of the above-gap photons in the photon-carrier generation process~\cite{CPL}. In general, those devices are promising for heat-to-electricity energy conversion in the common industry waste heat regime.

\section{Summary and outlooks}~\label{conclusion}

{\color{blue}
This paper attempts to provide a succinct review of the research frontier of inelastic thermoelectric effects. We summarized both theoretical and experimental progresses on inelastic thermoelectric transport and fluctuation in mesoscopic systems. We first give a general theoretical framework of the thermoelectric elastic and inelastic transport and revealed the unique role of the inelastic process of thermal transport in mesoscopic systems. We then show the distinct bounds on the linear transport coefficients of the elastic and inelastic thermoelectric transport from the general theoretical framework. We further summarize the unprecedented phenomena emerging from inelastic thermoelectric transport such as linear thermal transistor, cooling by heating, heat-charge cross rectification,  and cooling by thermal current. Inspired or based on inelastic thermoelectric effects, several approaches to improve thermoelectric performance are summarized, including heat-charge separation, thermoelectric cooperative effects, nonlinear enhancement of performance, non-canonical reservoirs, and near-field enhancement effect. For the near-field enhanced thermoelectric energy conversion, we discuss a set of examples including quantum-dots systems and graphene-h-BN-InSb systems.

Moreover, by integrating spin thermoelectric effect with the concepts from magnonics, the electron-magnon interactions for the nonequilibrium transport has been studied recently in many theoretical~\cite{Tulapurkar10,bauer12,chumak15,RezendePRB16,QaiumzadehPRL18,TangPRB18,UpadhyayPRE,WangPRL} and experimental works~\cite{uchida08,jaworski10,slachter10,walter11}. The asymmetric spin Seebeck effect, has recently been discovered both in metal/insulating magnet interfaces and magnon tunneling junctions, which leads to many interesting effects, such as spin thermal rectifiers effect~\cite{RenPRB13}, spin transistor effects~\cite{RenPRBTheory}, logic gates, and negative differential spin Seebeck effects~\cite{RenPRBPredicted}. These properties could have various implications in flexible thermal and spin information control. The generalization of the nanoscale metal-magnetic insulator interfaces with the electron-magnon interactions for inelastic thermoelectric transport and fluctuations in mesoscopic system is an interesting future direction.}

In addition to these contents reviewed above, there are still many interdisciplinary research frontiers of great curiosity, which are partially listed below:

(i) {\it Thermodynamic uncertainty relation}.
Recently, a thermodynamic uncertainty relation has been formulated for classical Markovian systems demonstrating trade-off between current fluctuation (precision) and dissipation (cost) in nonequilibrium steady state~\cite{TUR15,HasegawaPRL,TURPRE19,horowitz20,TURPRL22}.
The thermodynamic uncertainty relation implies that a precise thermodynamic process with little noise needs the high entropy production.
It is believed to be important in exploring the thermodynamic uncertainty bounds on the multi-terminal inelastic thermoelectric heat engine.

(ii) {\it Geometric-phase-induced pump.}
The second law of thermodynamics indicates that heat cannot be transferred spontaneously from low-temperature heat reservoir to high-temperature heat reservoir. To go beyond this conventional thermoelectric energy conversion,
a Berry-phase-like effect provides an additional geometric contribution~\cite{NemenmanPRL,Sinitsyn07,RenPRL10,WangPRA17,wangpump,wang2021,ArracheaPRXQuan,MyPRBGTUR} to pump electric and heat currents  against the thermodynamic bias.
Hence, it is intriguing to analyze the influence of the geometric-phase-induced pump
in the periodically driven quantum thermal machines~\cite{ArracheaPRB20}, e.g., the inelastic thermoelectric engine.

(iii) {\it Enhancing performance of NTPV systems via twisted bilayer two-dimensional materials.} The performance of NTPV systems can be greatly improved due to the hybridization effect of polaritons. Recently, the concept of photonic magic angles has attracted the attention of many researchers, due to the manipulation of the photonic dispersion of phonon polaritons in van der Waals bilayers~\cite{hu2020}. The twisted two-dimensional bilayer anisotropy materials or insulator slabs are explored in the near-field systems and the near-field radiative heat transfer can be significantly enhanced by the twist-nonresonant surface polaritons~\cite{HeOL20,TangACS,PengACS}. Inspired by this concept, it is extraordinarily promising to enhance the output power and energy efficiency of the NTPV systems by employing the twisted bilayer two-dimensional materials: the near-field absorber and emitter can be consisted of two dimensional anisotropic material/structure, e.g., Van der Waals materials or grating structures.

(iv) {\it Angular momentum radiation.}
The spin-orbit interaction, i.e., the coupling between the electron (light) orbital motion and the corresponding spin,
is fundamentally important in spintronics~\cite{awschalom09},
topological physics~\cite{HasanRMP,QiRMP},
and nano-optics~\cite{bliokh15,FrancoPR}.
Recently, the concept of angular momentum radiation was proposed via the spin-orbital interaction in the molecular junctions interacting with the electromagnetic waves~\cite{ZhangZQ20}. Based on the nonequilibrium Green's function method, the angular momentum selection rule for inelastic transport was unraveled.
Hence, it should be interesting in incorporating the angular momentum selection rule into the photon-involved inelastic thermoelectric machines.

\section{Disclosure statement}
No potential conflict of interest was reported by the authors.

\section{Funding}
This work was supported by the support from the funding for Distinguished Young Scienctist from the National Natural Science Foundation of China (Grant Nos. 12125504, 12074281, 12047541, 12074279, and 11704093), the Major Program of Natural Science Research of Jiangsu Higher Education Institutions (Grant No. 18KJA140003), the Jiangsu specially appointed professor funding, the Academic Program Development of Jiangsu Higher Education (PAPD), the Opening Project of Shanghai Key Laboratory of Special Artificial Microstructure Materials and Technology, the China Postdoctoral Science Foundation (Grant No. 2020M681376), the faculty start-up funding of Suzhou University of Science and Technology, and Jiangsu Key Disciplines of the Fourteenth Five-Year Plan (Grant No. 2021135).

\bibliography{Ref-transistor}

\end{document}